\documentclass[useAMS,usenatbib]{mn2e}
\usepackage{graphicx}
\usepackage{natbib}
\usepackage{epsfig}
\usepackage{amssymb}

\title[Large-scale jets from AGN as a source of ICM heating]{Large-scale jets from active galactic nuclei as a source of ICM heating: cavities and shocks}
\author[Perucho, Mart\'{\i}, Quilis and Ricciardelli]{Manel Perucho$^{1,2}$ \thanks{E-mail: manel.perucho@uv.es}, 
Jos\'e Mar\'{\i}a Mart\'{\i}$^{1,2}$, Vicent Quilis$^{1,2}$, Elena Ricciardelli$^{1}$\\
$^{1}$Departament d'Astronomia i Astrof\'{\i}sica. Universitat de Val\`encia. Av/ Vicent Andr\'es Estell\'es s/n, 46100 Burjassot (Val\`encia), Spain\\
$^{2}$Observatori Astron\`omic. Universitat de Val\`encia. C/ Catedr\`atic Jos\'e Beltr\'an, 2, 46980 Paterna (Val\`encia), Spain}

\begin{document}
\date{Released 2012 Xxxxx XX}

\pagerange{\pageref{firstpage}--\pageref{lastpage}} \pubyear{2012}

\label{firstpage}

\maketitle

\begin{abstract}
The evolution of powerful extragalactic jets is not only interesting by itself, but also for its impact on the evolution of the host galaxy and its surroundings. We have performed long-term axisymmetric numerical simulations of relativistic jets with different powers to study their evolution through an environment with a pressure and density gradient. Our results show key differences in the evolution of jets with different powers in terms of the spatial and temporal scales of energy deposition. According to our results, the observed morphology in X-ray cavities requires that an important fraction of the jet's energetic budget is in the form of internal energy. Thus, light, lepton-dominated jets are favoured. In all cases, heating is mainly produced by shocks. Cavity overpressure is sustained by an important population of thermal particles. Our simulations reproduce the cool-core structure in projected, luminosity-weighted temperature. We have performed an additional simulation of a slow, massive jet and discuss the differences with its relativistic counterparts. Important qualitative and quantitative differences are found between the non-relativistic and the relativistic jets. Our conclusions point towards a dual-mode of AGN kinetic feedback, depending on the jet power.
\end{abstract}


\section{Introduction} 
\label{intro}

     Extragalactic jets are usually observed in massive elliptical galaxies, either isolated or in rich clusters, and represent up to 1\% of the total number of known galaxies \citep{re12}. Radio-loud galaxies (with strong radio emission due to the presence of a jet) are most common among the more massive galaxies, representing up to 30\% of the galaxies with mass over $5\times10^{11}\,{\rm M_\odot}$ solar masses \citep{be05}. The jets are formed in the surroundings of supermassive black holes in (radio-loud) active galactic nuclei (AGN) as a result of magnetohydrodynamical processes \citep{bz77,bp82}. The jet is accelerated to relativistic velocities, as observed with the VLBI technique \citep[see, e.g.][]{li09}, and is thus supersonic. Consequently, it forms a strong shock in the ambient medium, which propagates at a fraction of the speed of light during the first kiloparsecs of evolution \citep{oc98}. The evolution of the jet after the first kiloparsec is related to its intrinsic properties (e.g., kinetic power) and those of the host galaxy (e.g., gas density). These shocks have been shown to propagate to kiloparsec scales in powerful radio sources, like Hercules~A
\citep{nu05}, Hydra~A \citep{si09b}, MS0735.6+7421 \citep{mc05}, HCG~62
\citep{git10}, 3C~444 \citep{cro11} or PKS B1358-113 \citep{sta14}, which is in agreement with the results reported in \citet{pqm11} (Paper~I, from now on) regarding the existence of large-scale, low-Mach number shocks around powerful radio sources. Nevertheless, shocks have been detected even in less-powerful jets \citep{kr07,cro07}, and this is also confirmed by numerical simulations \citep{pm07,bbrp11}.

   These jets present a morphological dichotomy between FRI and FRII type jets \citep{fr74}.
Whereas the former \citep[e.g., 3C~31,][]{lb02} show disrupted structure consisting of plumed lobes at
kiloparsec scales, the latter \citep[e.g., Cyg A,][]{cb96} are
highly collimated and show bright hotspots at the interaction region with the ambient medium. Even though the Lorentz factors of both FRI and FRII jets appear to be similar at the parsec scales \citep{gi01,cg08}, the
present paradigm of FRI jet evolution \citep{bi84,la93,la96} states that these jets are decelerated at kiloparsec scales due to entrainment of gas and their properties become different from those of FRII jets. The origin of this dichotomy is attached to a complex
combination of several intrinsic (e.g., jet power) and external (i.e.,
environmental) factors. The main processes invoked to explain this difference are entrainment 
\citep{dy86,dy93,bi94} and mass load by stellar winds \citep{ko94,BL96,lb02,HB06}. Hybrid morphology sources, in which
one of the jets shows FRII morphology and the other jet shows FRI morphology \citep{gw00}, 
have been invoked as evidence for the role of the ambient medium in this dichotomy, but the different effect of the growth of instabilities \citep[e.g.,][]{pe10}, the mass load by stellar winds \citep{BL96,pmlh14} or the clear correlation between the jet power and its morphology \citep{rs91} rule it out as the only important parameter in the dichotomy. 
It has been recently suggested that there could be a transition in time between FRII and FRI morphology, after a decrease in the jet injection power \citep{wa11}. However, this hypothesis should explain why FRI jets show relativistic speeds at the parsec scales \citep[see, e.g.][]{gi01}.    

  A cornerstone question in galactic Astrophysics and Cosmology is the interplay between galactic activity and the evolution of the host galaxy, via changes in the interstellar medium and in the evolution of the supermassive black hole at its nucleus, or with the X-ray emitting intracluster gas (ICM). The ICM emits thermal (bremsstrahlung) radiation and cools down at a rate that implies short cooling times as compared to the age of the Universe. After losing energy, the gas should fall onto the galactic potential well. The expected falling rate of such cooling flows is $10\,-\,100\,{\rm M_\odot\,yr^{-1}}$. In the last decades, the observations of clusters of galaxies at the X-ray band with the space observatories \emph{XMM-Newton} and \emph{Chandra} revealed a lack of cool gas in the centers of many galaxies, contrary to expectations \citep[see, e.g.,][and references therein]{mn07,fb12}. 

This cooling flow problem has also
important implications regarding the star formation rates in those galaxies and the growth of the central black hole \citep[see,
e.g.,][]{mn07,fb12,cat09}. In fact,  galaxy formation models neglecting the AGN heating lead
to an overproduction of stars in massive galaxies \citep{oser10,
lackner12, navarro13} which results in galaxies overly massive and star
forming. This overcooling problem appears much alleviated in
hydrodynamical simulations which implement self-consistent subgrid
models of AGN  feedback (e.g. \citealt{sijacki07, dub10, dub13, Vogel14}).

    Thus, the cooling flow problem is an issue that has triggered a large list of publications including new observations or theoretical attempts to explain it. The most accepted heating mechanisms proposed to stop the cooling flows are related to galactic activity \citep[see, e.g.,][and references therein]{mn07,fb12}. The observed anti-correlation between the radio lobes formed by jets and the X-ray emission from the cluster gas \citep[e.g.,][]{mn07,fa06} raised the idea that the buoyancy of under-dense cavities formed by these jets could perform mechanical work on the ambient gas by displacing it and thereby heating it by compression. Moreover, significant levels of metallicity have been recently detected at considerable distances from the active galaxy, which is possible only if an outflow dragged those metals produced in the stars within the galaxy and brought them away \citep{si09a,kir09,we10,we11,kir11}. The amount of work required to displace the ambient gas from the cavities is computed using simple arguments \citep[$W\,=\,p\,V$, with $p$ the pressure measured for the ambient gas and $V$ an estimate of the volume of the X-ray cavity, e.g.][]{mc05}. The ages of those bubbles can be inferred using an estimate for the velocity of buoyant motion of the cavity. This information and the estimate of the work done to inflate the X-ray cavity, results in values for the AGN output power \citep{bir08} that are compatible with typical jet powers \citep[see, e.g.,][]{wil99,gc01a,lcp12}. However, recent observations have shown that the lobes are surrounded by shocks with low Mach numbers \citep[$M_{\rm s}\,=\,1\,-\,2$][]{mc05,nu05,si09b,git10} in powerful sources, i.e., they have not reached the buoyancy stage, but evolve owing to pressure difference with the environment. This would imply faster evolution (pressure driven) and, accordingly, larger jet powers than estimated. 
 
   A large amount of theoretical work has been performed, mainly via numerical simulations, attempting to explain the process of heating and, more recently, feedback. Heating by the buoyant bubbles formed by relic radio lobes has been extensively studied \citep{ch01,qb01,br02,ch02,rb02,bi04,dv04,ry04,br06,ro07,ss08,ss09,dy10}. In this scenario the heating process occurs by mixing after the development of Rayleigh-Taylor or Kelvin-Helmoltz instabilities in the boundary between the bubble and the ICM, and in the turbulent wake of the buoyant motion. The mixing produces a net gain of internal energy of the ICM that has been claimed to be efficient enough to stop or delay the cooling flows. Another approach to this problem consisted on the injection of a jet (or mass and energy) into a numerical grid filled by the ambient medium \citep[e.g.,][]{ch01,qb01,br02,rey02,om04,ob04,za05,br06,vr06,ct07,vr07,br07,bin07,bsh09,oj10,gas11a,gas11b,ga12}. Many of the jet simulations do not take into account the relativistic nature of the jet flow and thus require unrealistic initial jet radii or large mass flows. The justification is the observation of massive, slow flows in a number of active galaxies and that most clusters seem to preserve the cool-core structure and do not invert their temperature gradients \citep{mit09}. This requires a local and gentle mechanism \citep[e.g.][]{gas11a,gas11b}. However, massive, slow flows are only observed at distances that range from several kiloparsecs to tens of kiloparsecs at most \citep[e.g.,][]{mo05,mo07,ho08,nes08,gui12,mo13} and appear typically associated with a faster radio-jet with much larger sizes, so they could well be a consequence of the action of the relativistic jet on the ambient medium, more than a main actor of the whole process \citep[e.g.,][]{nes08,gui12,mo13}. 
   
  The simulations of buoyant bubbles are all based on observational evidence of pressure equilibrium between the X-ray cavity and the cluster gas, and pressure-driven evolution is neglected. Different measures of lobe pressures similar or even smaller than the pressure of the surrounding gas support this idea \citep{cro04,cro05}. These measures show that a dominant population of relativistic protons is unlikely \citep{cro05}. However, in the published calculations of the lobe pressure, the possible thermal component of the lobe gas is neglected because it is difficult to estimate. This component could be even the dominant population in the lobes, so the values obtained would be severely underestimated.
Pressure equilibrium with the ambient medium could be thus reached in very old or low power jets, but it is not the case in active or powerful jets, as shown by the detections of shocks around their lobes \citep{mc05,nu05,si09b,git10,sta14}. As recently shown in Paper~I, and by \citet{wb11}, shocks may be extremely important and efficient in the heating process of the galactic and cluster gas, and can displace large amounts of gas from the host galaxy, thus quenching star formation. 

   \citet{sha11} have shown, via a study of the colour evolution of local galaxies, that there is a dual mode AGN kinetic feedback, which can be divided into FRI and FRII-like feedback. The latter is claimed to be extremely relevant for the evolution of the host galaxy and its environment. A possible positive feedback effect consisting of star formation triggered at the regions of shocked ambient medium has been claimed by \cite{ga12}. This work shows that the star-forming process would be the response to the compression of clouds in the ISM when the jet still propagates within the host galaxy. Although a large amount of work has been done to study the different possible mechanisms and their effects, we still do not have a clear idea of the relevant processes in different known scenarios or their relative importance. For instance, the role of the magnetic fields in thermal conduction or viscous dissipation remains unclear.  

  In Paper~I we presented our first analysis on the influence of powerful relativistic jets on their environment, mainly driven by the efficiency of conversion of the injected energy into ambient-medium heating by a strong shock, on the basis of  four simulations of relativistic jets, differing in jet power and composition. The analysis was focused on one of these simulations, in which the shock-heating was clearly observed. In addition, the evolution of the different simulations was modeled using an analytical approximation. In this work we analyze in detail the whole set of simulations, paying attention to the similarities and differences among them in different terms. In particular, we discuss the differences in their effect on the ambient medium and their morphologies. We also add the results from a Newtonian (non-relativistic) simulation that is compared with its relativistic counterpart. These simulations represent the longest timespan yet produced for relativistic hydrodynamics simulations of jets with different powers and compositions, with the aim to study 1) the interaction with the ambient medium and its effects, and 2) the long-term properties of jets and the cocoons that they form and are observed as radio-lobes or X-ray cavities. For this, we used two-dimensional axisymmetric simulations with the code \emph{Ratpenat}, performed in the supercomputer Mare Nostrum, with a total amount of $\simeq 10^6$ computing hours. This is by far, the largest computational effort done to study the long-term evolution of extragalactic jets. The paper is structured as follows: the setup of the simulations, together with the parameters used are presented in Section~2, and the results are given in Section~3. Section~4 includes a discussion of the results. A summary and the conclusions of this work are provided in Section~5.

\section{Simulations}
\label{s:sim}

\subsection{Computational setup}
\label{ss:setup}
The simulations presented in this paper use the finite-volume code {\it Ratpenat} \citep{pe10}. {\it Ratpenat} is a hybrid -- MPI + OpenMP -- parallel code that solves the equations of relativistic hydrodynamics in conservation form using high-resolution-shock-capturing methods \citep[see][and references therein]{pe10}: i) primitive variables within numerical cells are reconstructed using PPM routines, ii) numerical fluxes across cell interfaces are computed with Marquina flux formula, iii) advance in time is performed with third order TVD-preserving Runge-Kutta methods.

The numerical grid is structured as follows: in the radial direction, a grid
with the finest resolution extends up to 50~kpc (Model J45l, J1 in Paper~I) or 100~kpc (Models
J46 -J2 in Paper~I-, J44 -J3 in Paper~I-, J45b -J4 in Paper~I-). An extended grid with decreasing resolution was added 
up to 1~Mpc. Along the axis, the grid extends up to distances close to 1~Mpc with 
homogeneous resolution (50-100~pc/cell). This translates into a total grid size of around
$1800 \times 10000$ (radial and axial, respectively) cells. The time-step during the first part of the simulations, when the
jet is still active, was 50 to 100 years. The boundary conditions in the simulations are reflection at the jet base, to mimic the presence of a
counter-jet, reflection at the jet axis and outflow at the end of the grid in the axial and radial directions.

 The equations that are solved by the code are those corresponding to the conservation of mass,
momentum and energy. These conservation equations are, in the case of a relativistic flow
in two-dimensional cylindrical coordinates ($R,\, z$), assuming
axisymmetry and using units in which $c=1$:
\begin{equation}
  \frac{\partial \mathbf{U}}{\partial t} + \frac{1}{R}\frac{\partial R
\mathbf{F}^R}{\partial R} + \frac{\partial \mathbf{F}^z}{\partial z} =
\mathbf{S} ,
\end{equation}
with the vector of unknowns
\begin{equation}
  \mathbf{U}=(D,D_l,S^R,S^z,\tau)^T ,
\end{equation}
fluxes
\begin{equation}
  \mathbf{F}^R=(D v^R , D_l v^R , S^R v^R + p , S^z v^R , S^R - D v^R)^T ,
\end{equation}
\begin{equation}
  \mathbf{F}^z=(D v^z , D_l v^z , S^R v^z , S^z v^z + p, S^z -D v^z)^T ,
\end{equation}
and source terms
\begin{equation}
  \mathbf{S}  =  (0, 0, p/R + g^R, g^z, v^R g^R + v^z g^z)^T .
\end{equation}
 
 The five unknowns $D,D_l,S^R,S^z$ and $\tau$, refer to the densities of
five conserved quantities, namely the total and leptonic rest masses,
the radial and axial components of the momentum, and the energy
(excluding rest-mass energy). They are all measured in the laboratory
frame, and are related to the quantities in the local rest frame of the
fluid (primitive variables) according to
\begin{equation}
  D = \rho W,
\end{equation}
\begin{equation}
  D_l = \rho_l W,
\end{equation}
\begin{equation}
  S^{R,z} = \rho h W^2 v^{R,z},
\end{equation}
\begin{equation}
  \tau=\rho h W^2\,-\,p\,-\,D,
\end{equation}
where $\rho$ and $\rho_l$ are the total and the leptonic rest-mass
densities, respectively, $v^{R, z}$ are the components of the velocity
of the fluid, W is the Lorentz factor ($W = 1/\sqrt{1-v^i v_i}$, where
summation over repeated indices is implied), and $h$ is the specific
enthalpy defined as
\begin{equation}
  h = 1 + \varepsilon + p/\rho,
\end{equation}
where $\varepsilon$ is the specific internal energy and $p$ is the
pressure. Quantities $g^R$ and $g^z$ in the definition of the
source-term vector ${\bf S}$, are the components of an external gravity
force that keeps the atmosphere in equilibrium (see Sect.~\ref{ss:ambient}).

The system is closed by means of the Synge equation of state
\citep[][described in Appendix A of Perucho \& Mart\'{\i} 2007]{sy57} that accounts for a mixture
of relativistic Boltzmann gases (in our case, electrons, positrons and
protons). The code also integrates an equation for the jet mass fraction, $f$. This quantity, set to 1 for the injected beam material and 0 otherwise, is used as a tracer of the jet material through the grid. In these simulations, cooling has not been taken into account, as the typical cooling times in the
environment are long compared to the simulation times \citep[see Figure 10 in][]{hr02}.

The simulations were performed in Mare Nostrum, at the
Barcelona Supercomputing Center, and Tirant, at the University of Val\`encia, within the \emph{Red Espa\~nola de Supercomputaci\'on} (Spanish Supercomputing Network), with up to 128 processors, which were added as the jet evolved (starting
with 16 processors). These simulations required around $2 \times 10^5$ computational hours, depending on the model,
resulting in a total of around $10^6$ hours.

\subsection{Jet parameters}
\label{ss:jets}  
 
   We performed four simulations of 2D axisymmetric, relativistic jets with the aim to study the global structure and dynamics of the cocoon-shocked ambient medium system depending on the jet power and composition. The detailed list of parameters used in each simulation is listed in Table~\ref{tab1}. These jets, with kinetic powers ranging between $L_{\rm k}=10^{44}$ and $10^{46}\,{\rm erg\,s^{-1}}$, are injected at 1~kpc, with a radius, $R_{\rm j}=100$~pc. The flow velocities at injection range from $v_{\rm j}=0.9\,c$ to $v_{\rm j}=0.99\,c$, and typical density ratio between the jet material and environment of $\rho_{\rm j}/\rho_{\rm a} = 5\times 10^{-4}$. The injection of the jets with these characteristics lasted for 16-50~Myrs, depending on the model. After this time the jet injection velocity is continuously reduced down to zero\footnote{At the switch-off time, the injection velocity starts to decrease following a power law which reduces it to zero in a typical time scale of $10$~Myrs.}, implying no injection, and we still follow the evolution of the whole system (Paper~I). The resolution is 1 cell per jet radius at injection in models J46 and J44, and 2 cells per jet radius in the case of J45l. This small resolution is justified by two main reasons: i) as the jet expands along the grid, it is progressively resolved by a larger amount of cells, with 12 (24 in the case of J45l) cells/$R_{\rm j}$ at 100~kpc and 32 (64 in the case of J45l) at 400~kpc, and ii) in this work we want to focus more on the overall jet-cavity-shocked ambient medium structure, so the resolution within the jet is not a crucial point. 
 
  An additional simulation of a wide and slow, non-relativistic jet was performed in order to have a direct comparison of its effects on the ambient medium as compared to the relativistic case. The parameters were chosen to obtain the same injection power as in J46. We used a large flow radius at injection, $R_{\rm j}=3\,{\rm kpc}$, injection velocity, $v_{\rm j}=0.3\,c$, and electron-proton composition (cf. the leptonic composition of J46). With these numbers, the jet temperature is $3.3\times 10^8 \,{\rm K}$ and the jet internal Mach number is $18.2$. This is a clearly unrealistic parameter set for extragalactic, powerful outflows, if we consider that the injection of the jet is located at $1\,{\rm kpc}$ from the active nucleus and the jet radius is 3 times larger, and also the relativistic nature of those jets \citep{bri94}. However, these are the numbers needed to obtain an outflow with a kinetic power of $10^{46}\,{\rm erg/s}$, and typically used in the literature to simulate those massive and slow outflows. An alternative is to increase the jet mass flux by increasing the density. In this case, keeping $v_{\rm j}=0.3\,c$ we obtain ballistic propagation and a very thin cocoon, which is in disagreement with observed morphologies of extragalactic jets.

%
\begin{table*}
  \begin{center}
  \caption{Parameters of the simulated jets. From left to right the columns give the model, the injection velocity, the
injection density, the leptonic number fraction, the jet radius at injection, the jet power, the maximum resolution,
and the switch-off time. Please note that the names have been changed with respect to those in Paper~I, with J45l corresponding to J1, J46 to J2, J44 to J3, and J45b to J4 in that paper. 
}
  \label{tab1}
 {\small
  \begin{tabular}{|l|c|c|c|c|c|c|c|}\hline 
{\bf Model} & {\bf v}$_{\rm j}$ [$c$] & {\bf $\rho_{\rm j}$} [g/cm$^3$] & \emph{\bf X}$_{\rm e}$ &
\emph{\bf R}$_{\rm j}$ [pc]  & \emph{\bf L}$_{\rm k}$ [erg/s] & 
{\bf max. resol.} [pc/cell] & {\bf t}$_{\rm off}$ [Myrs]  \\  \hline
J44 &  0.9 & $8.3\times 10^{-30}$ & 1.0 & $10^2$& $10^{44}$ & 100 & 50\\ 
J45l &  0.9 & $8.3\times 10^{-29}$ & 1.0 & $10^2$& $10^{45}$ & 50 & 50 \\
J45b &  0.9 & $8.3\times 10^{-29}$ & 0.5 & $10^2$& $10^{45}$ & 100 & 50 \\ 
J46 &  0.984 & $8.3\times 10^{-29}$ & 1.0 & $10^2$ & $10^{46}$ & 100 & 16\\ 
J46n & 0.3 &  $8.3\times 10^{-29}$ & 0.5 & $3\times 10^3$& $10^{46}$ & 100 & 16\\
\hline
  \end{tabular}
  }
 \end{center}
\end{table*}
%
\subsection{Ambient medium}
\label{ss:ambient}

%
\begin{figure*}
  \includegraphics[width=0.45\textwidth]{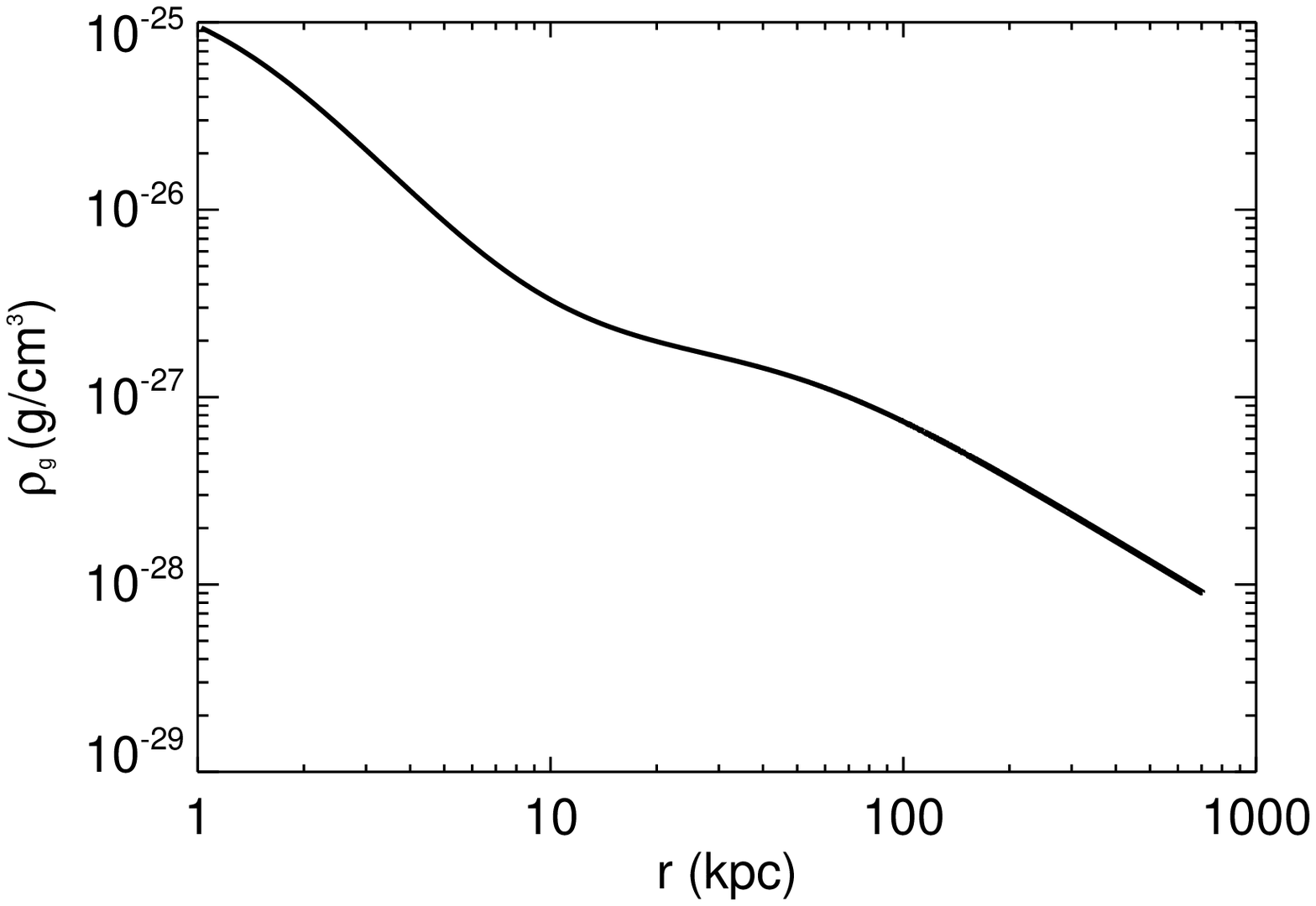}
 \includegraphics[width=0.45\textwidth]{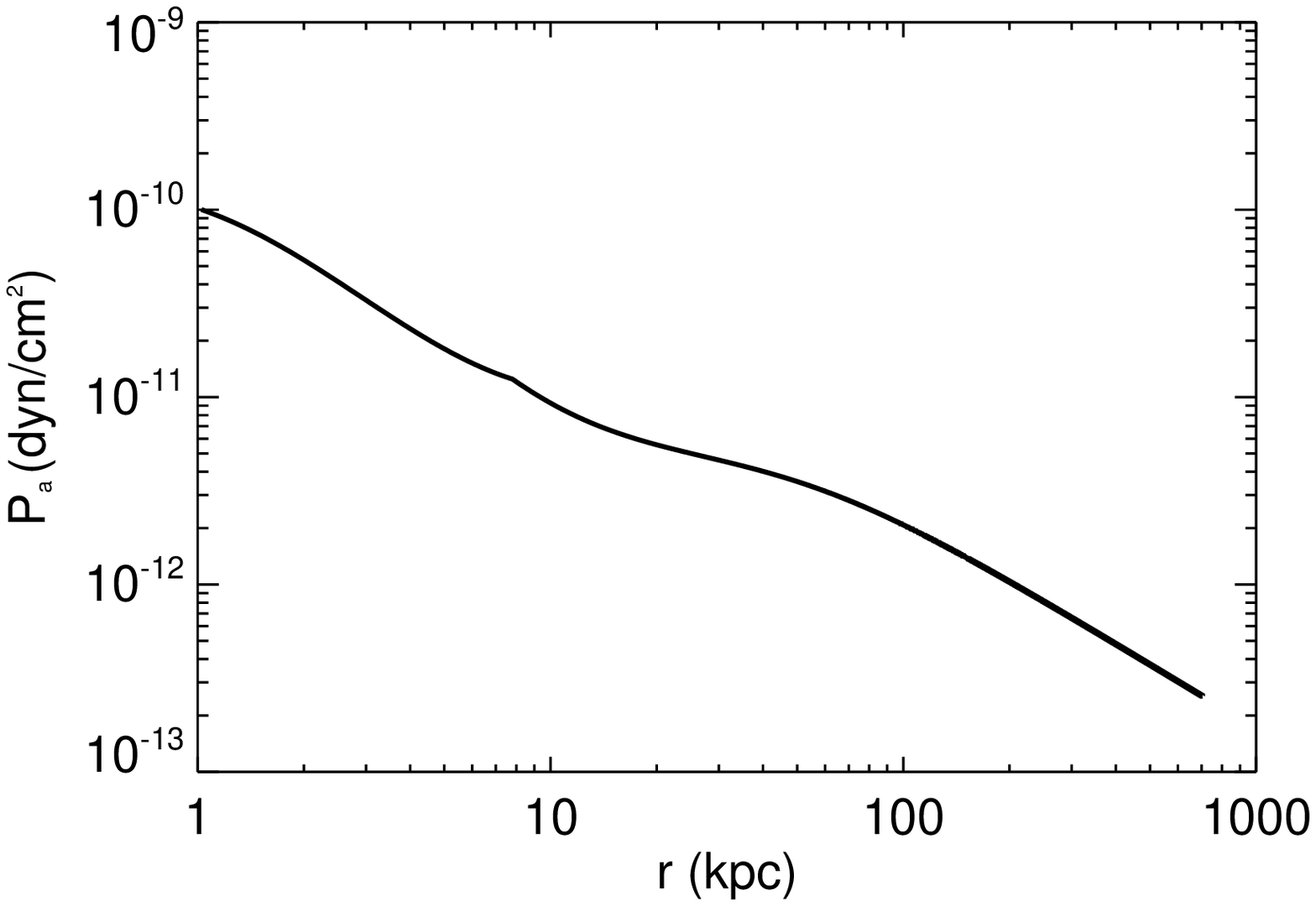}
 \caption{Density and pressure profiles of the ambient medium (cgs units) as defined in Eqs.~(\ref{next}) and (\ref{pext}).}
 \label{fig:ambient}
 \end{figure*}
%

The jets are injected in a
computational domain filled with an ambient medium in hydrostatic equilibrium 
with a King-like density profile that takes into account the elliptical galaxy -- origin of the jet -- and the galaxy cluster. 
The density profile parameters follows from fits to the X-ray data of the source 3C~31 \citep{ko99,hr02}.
The profile for the number density of such a medium is \citep{hr02,pm07}:
\begin{eqnarray}\label{next}
  n_{\rm ext} = n_{\rm c} \left(1 +
\left(\frac{r}{r_{\rm c}}\right)^2\right)^{-3\beta_{\rm atm,c}/2} +  \nonumber \\
+ n_{\rm g} \left(1 + \left(\frac{r}{r_{\rm g}}\right)^2\right)^{-3\beta_{\rm atm,g}/2},
\end{eqnarray}
with $n_{\rm c} = 0.18$~cm$^{-3}$, $r_{\rm c} = 1.2$~kpc,  $\beta_{\rm atm,c} =
0.73$, $n_{\rm g} =0.0019$~cm$^{-3}$, $r_{\rm g} = 52$~kpc, and  $\beta_{\rm atm,g} =
0.38$. The dark matter halo accounting for the external gravity can be fitted by a NFW density profile
\citep{nfw97}. All these parameters represent a moderate size galaxy cluster
with mass $10^{14}\,M_{\odot}$ and $\sim 1\, \rm{Mpc}$ virial radius. The corresponding temperature profile is \citep{pm07,hr02}:
\begin{eqnarray}\label{text}
  T_{\rm ext} = T_{\rm c} + (T_{\rm g} - T_{\rm c}) \frac{r}{r_{\rm m}} \,\,\,\,\,\, {\rm for\,} r\leq r_{\rm m} \nonumber \\
  T_{\rm ext} = T_{\rm g} \,\,\,\,\,\,  {\rm for\,} r > r_{\rm m}
\end{eqnarray}
where $T_{\rm c}$ and $T_{\rm g}$ are characteristic temperatures of the host galaxy
and the group ($4.9 \times 10^6$ K and $1.7 \times 10^7$ K,
respectively, and $r_{\rm m} = 7.8$ kpc is the matching radius. The external
pressure is derived from the number density and temperature profiles
assuming a hydrogen perfect gas \citep{hr02,pm07}:
\begin{equation}\label{pext}
  p_{\rm ext} = \frac{k_{\rm B} T_{\rm ext}}{\mu X} n_{\rm ext},
\end{equation}
where $\mu= 0.5$ is the mass per particle in amu, $X=1$ is the abundance
of hydrogen per mass unit, and $k_{\rm B}$ is the Boltzmann's constant. The pressure and density profiles in the (spherical) radial 
direction are plotted in Fig.~\ref{fig:ambient}.

\section{Results}
\label{s:res}

  The dynamics of the system is dominated by the jet active phase, whose
propagation through the ambient medium generates a characteristic
morphology: i) a bow-shock that acts on the ambient medium, ii) a
terminal or reverse shock at the head of the jet where the flow
decelerates and heats (forming the so-called hot-spot) and iii) the
cocoon inflated by the shocked jet particles and polluted with shocked
ambient medium via instabilities arising at the contact discontinuity
between both media. This cocoon is typically hotter and underdense compared with the
ambient medium. After the switching off the jet head
velocities quickly drop because of the short time scales needed by the relativistic flow to reach the jet terminal shock. 
Then, the bow-shock approaches sphericity very fast.

\subsection{Active phase}
\label{s:ap}

The evolution of the cocoon-shocked ambient medium system for model J46 was studied in Paper~I, although this evolution is qualitatively similar in all the models. The evolution can be divided into three main
phases: one-dimensional, two-dimensional and Sedov. The first two phases
correspond to the active phase of the jet, whereas the last one corresponds to 
the passive phase once the jet has been switched-off. In this section we 
describe and compare the active phase for models J44, J45l, J45b and J46.

\begin{figure}
 \includegraphics[width=\columnwidth]{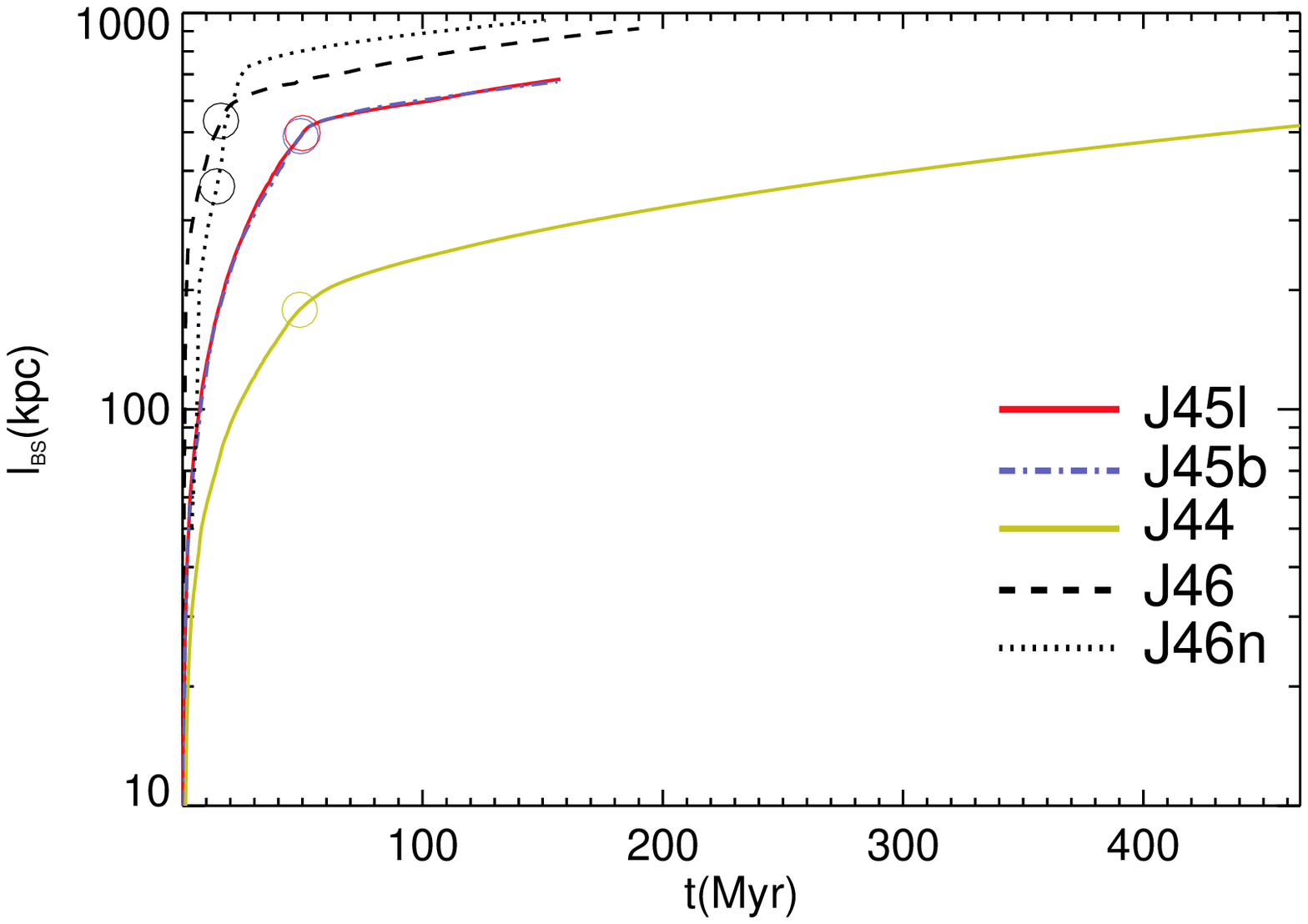}\\
 \includegraphics[width=\columnwidth]{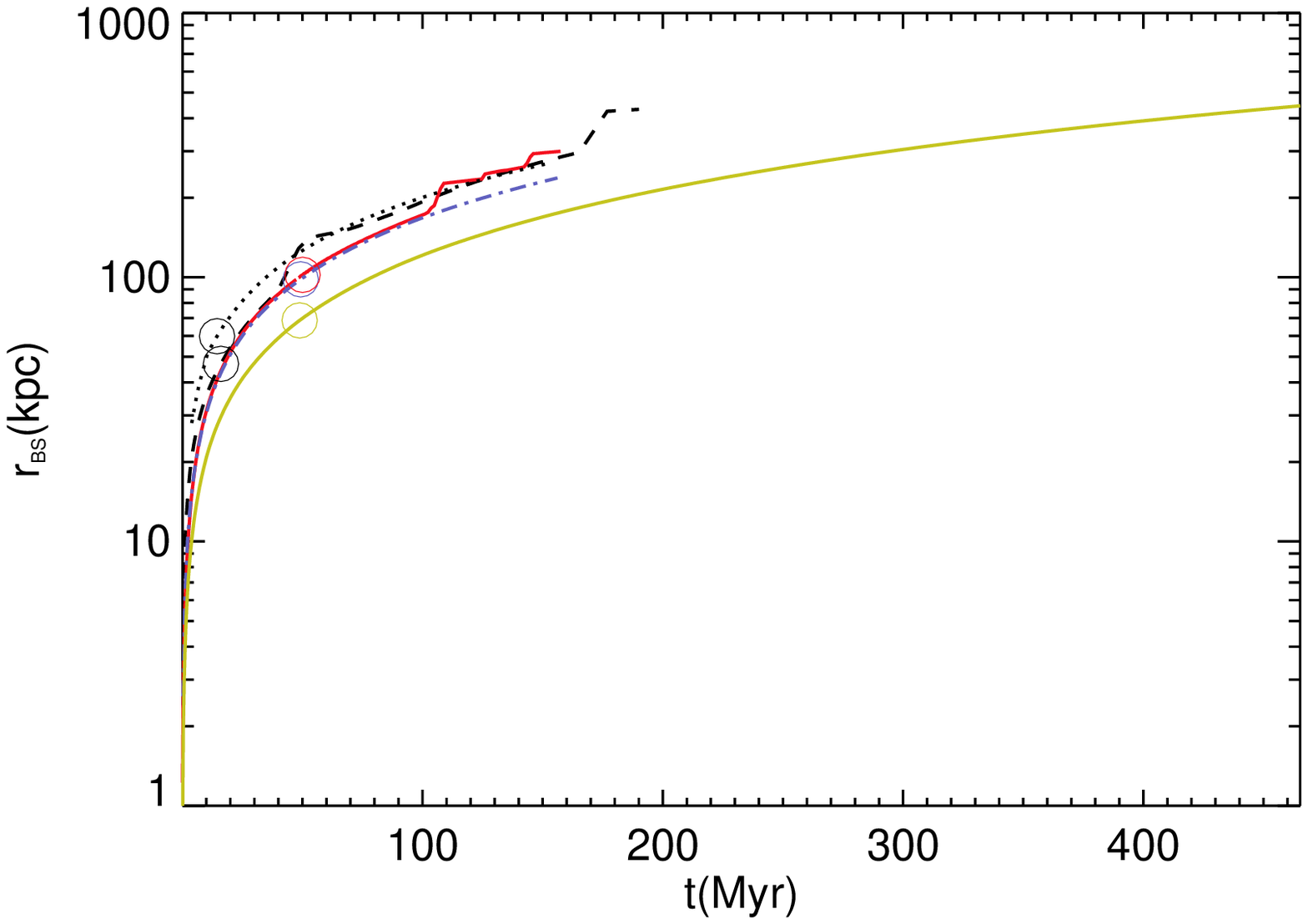}\\ 
  \includegraphics[width=\columnwidth]{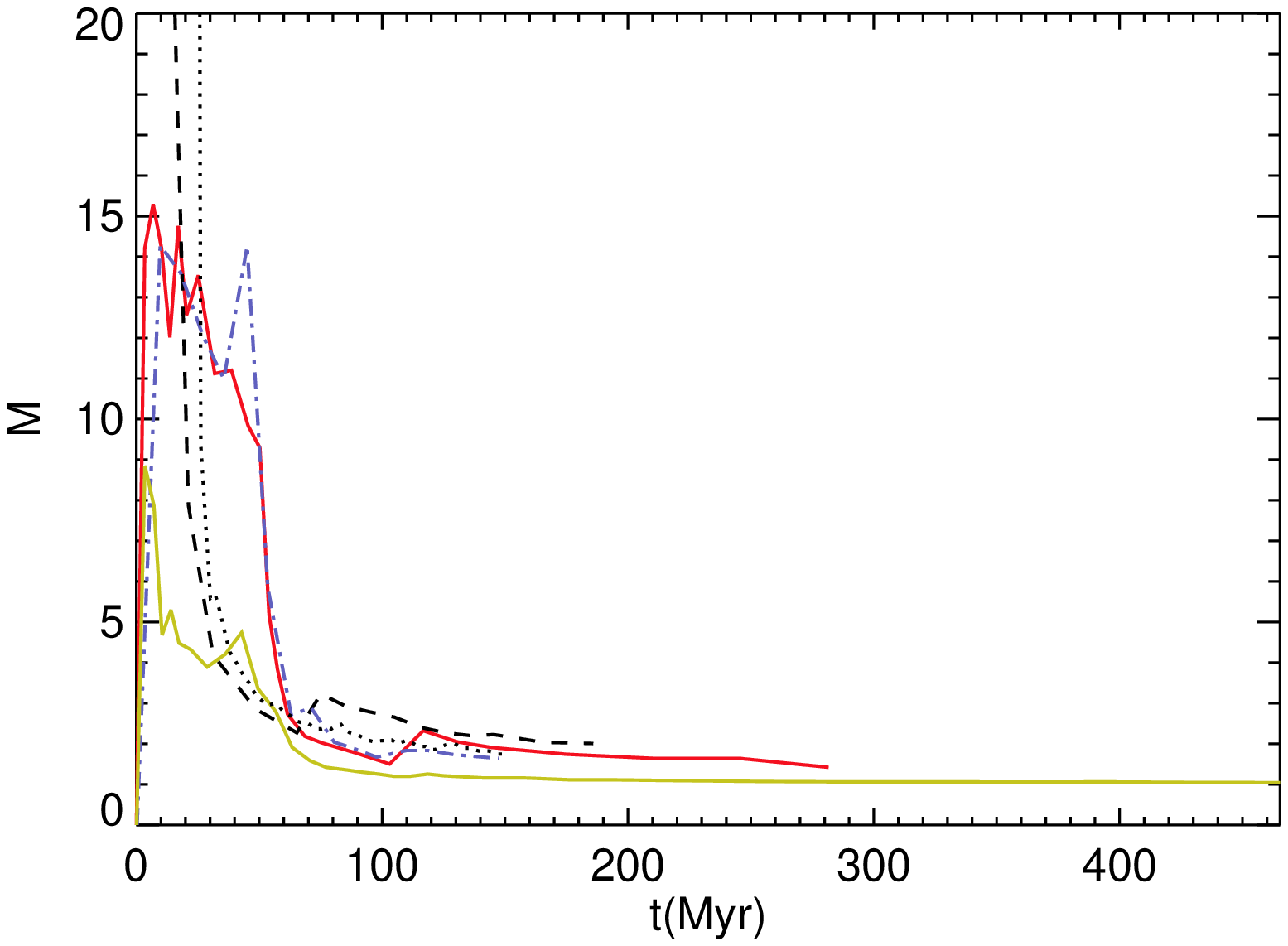}
 
 \caption{Position of the head of the bow-shock along the axis of the
jet (top panel), mean bow-shock radius (central panel), and Mach number
(bottom panel) versus time. The colours indicate the different models: solid green is J44, solid red is J45l,
dash-dotted blue is J45b, dashed black is J46, and dotted black is J46n. The position of the shock is defined as
that of the first numerical zone from the grid boundary with a non-zero
flow velocity. The Mach number has been obtained using a constant ambient-medium sound speed (as
corresponds to its isothermal nature at $r \geq 7\,$kpc) of $2.28\times
10^{-3}\,c$. The lines have been smoothed for the sake of clarity. 
The end of the active phase is indicated by circles.}
 \label{fig:lr}
 \end{figure}

 In a homogeneous ambient medium, during the so-called
one-dimensional phase, the jet propagates at its estimated
one-dimensional speed \citep{ma97}. In the present simulations, the
evolution in a density decreasing atmosphere accelerates the jet
propagation speed beyond the one-dimensional estimate. The onset of
the multidimensional phase is triggered by multidimensional, dynamical
processes taking place close to the jet's head. The change between the
one-dimensional and two-dimensional phases occurs in a few ($1.5 -
2.6$) million years for all the models. During the multidimensional
phase, the head of the jet decelerates increasing the flux of jet 
material into the cocoon. This increased flux of material plus the density profile of 
the ambient medium results in a fast expansion of the cocoon, which pushes the 
bow-shock. Figures~\ref{fig:lr} shows the
position of the head of the bow shock along the axis and the mean radius
in the transversal direction. The end of the two-dimensional phase (and
the beggining of the Sedov phase) for the different models are
indicated with circles in the figure. During the two-dimensional
phase, the bow-shock expands in the axial direction with velocities
$0.01\,-0.06\,c$ ($0.01\,-0.03\,c$, in the case of model J44). The
corresponding Mach numbers range between $5$ and $30$ ($3$ and $10$,
in the case of model J44, see Fig.~\ref{fig:lr}). 

%
%

\begin{figure*}
 \includegraphics[width=0.45\textwidth]{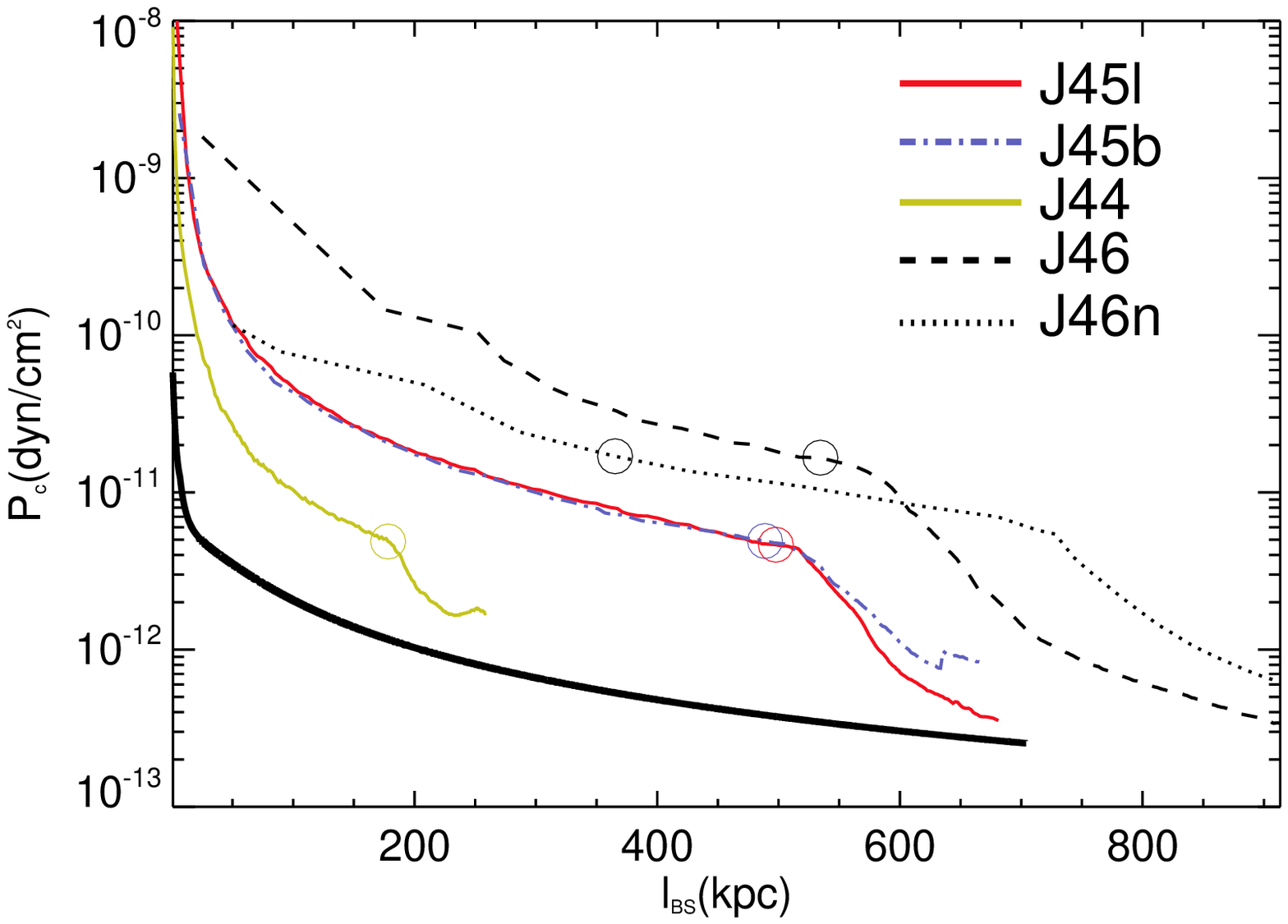}
 \includegraphics[width=0.45\textwidth]{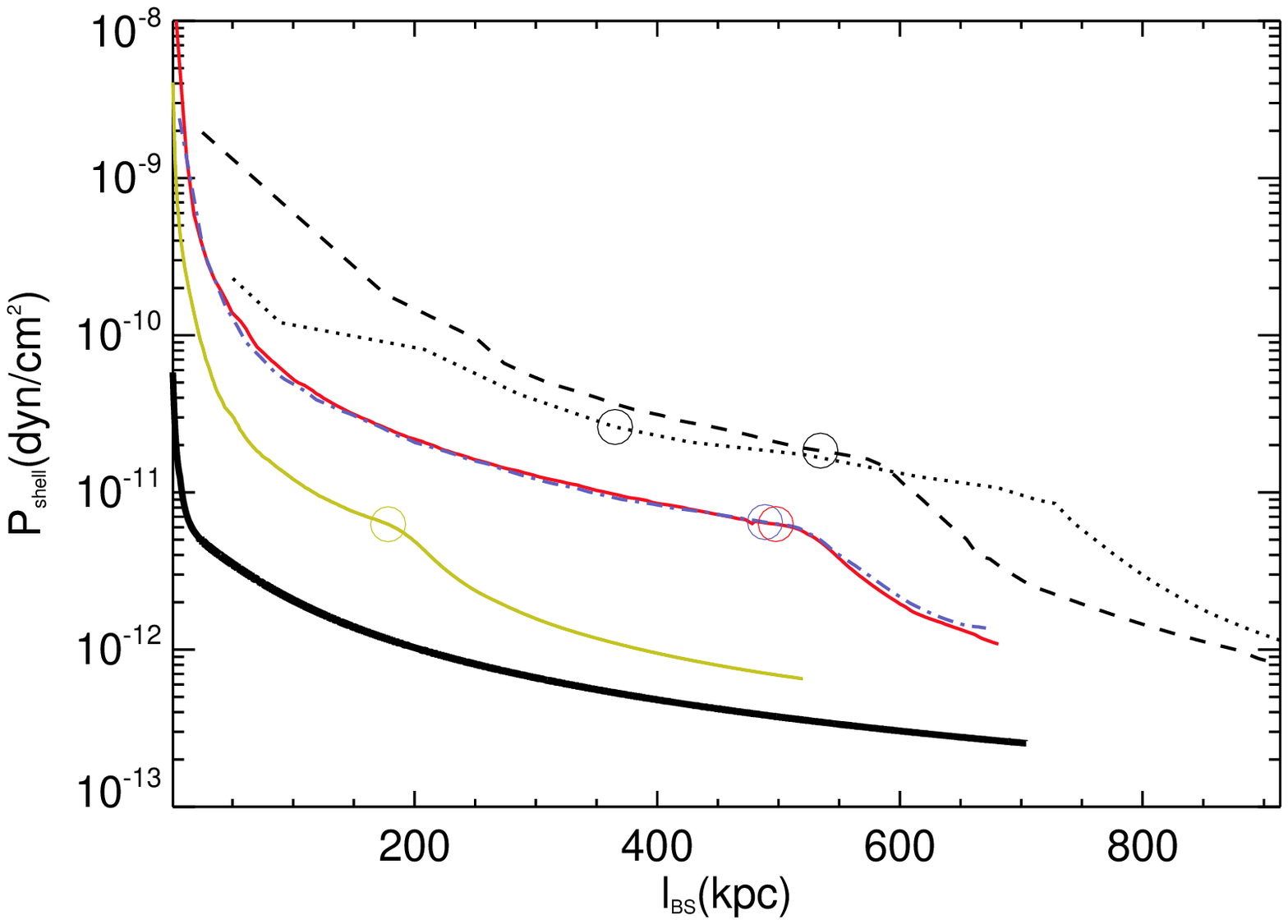} \\
  \includegraphics[width=0.45\textwidth]{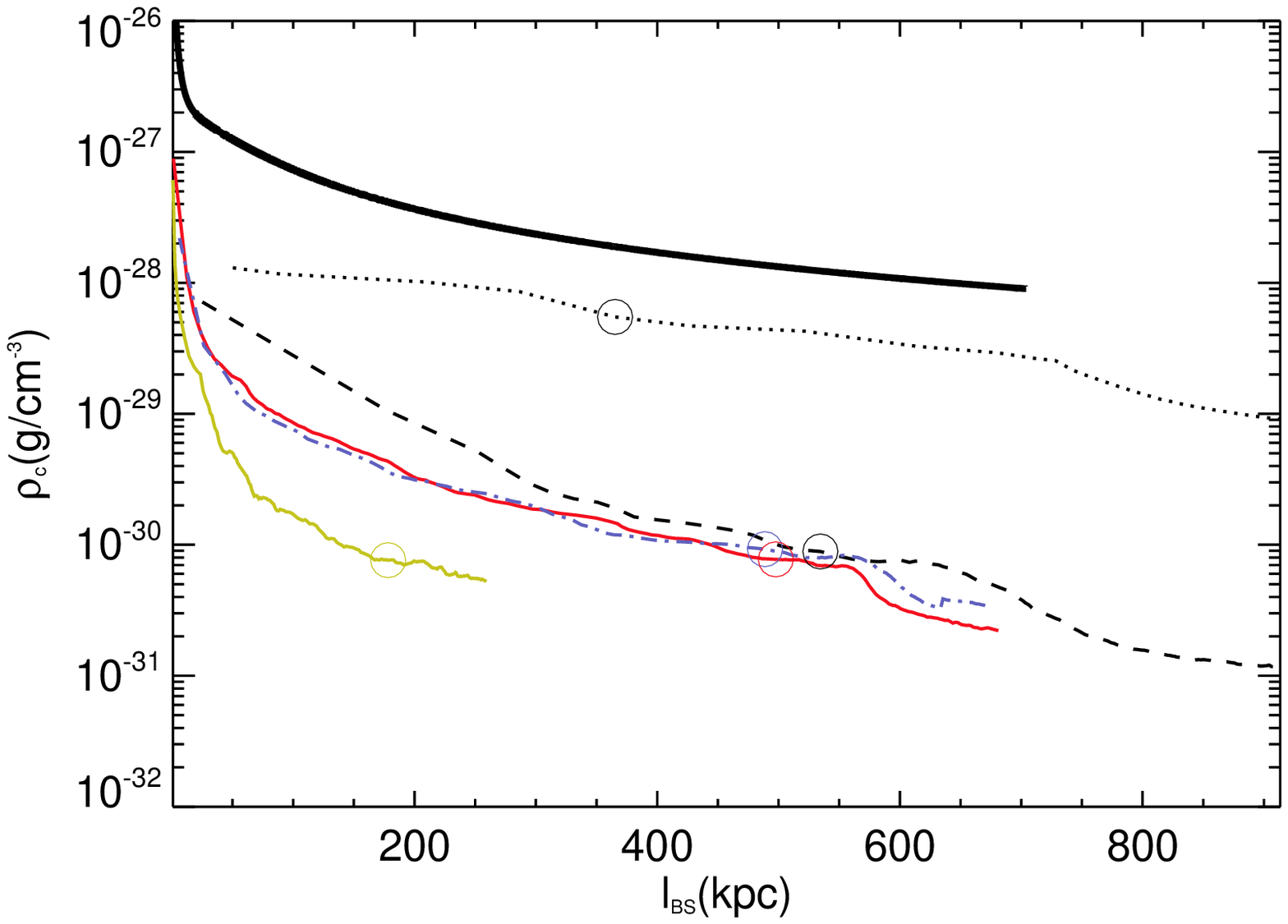}
 \includegraphics[width=0.45\textwidth]{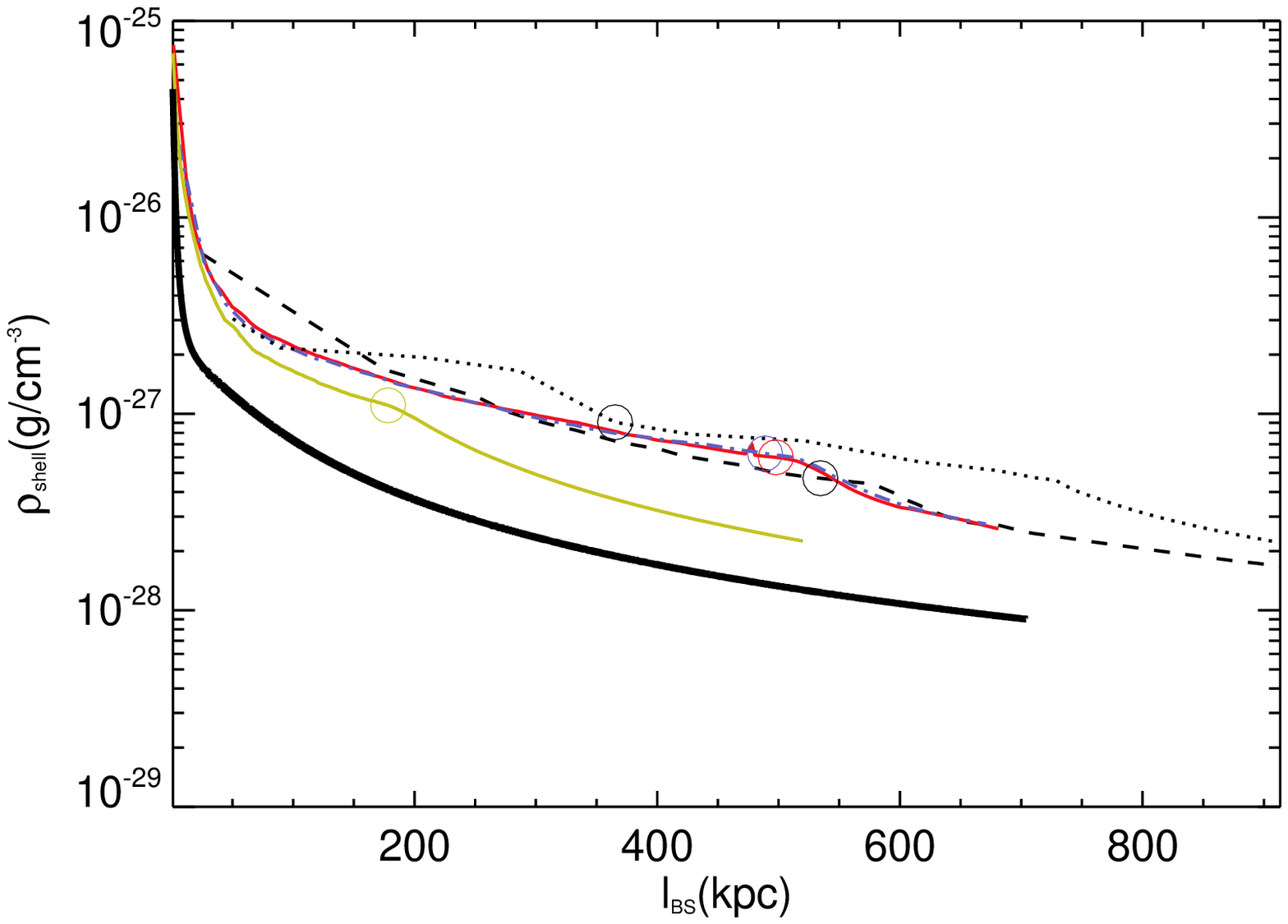}\\
 \includegraphics[width=0.45\textwidth]{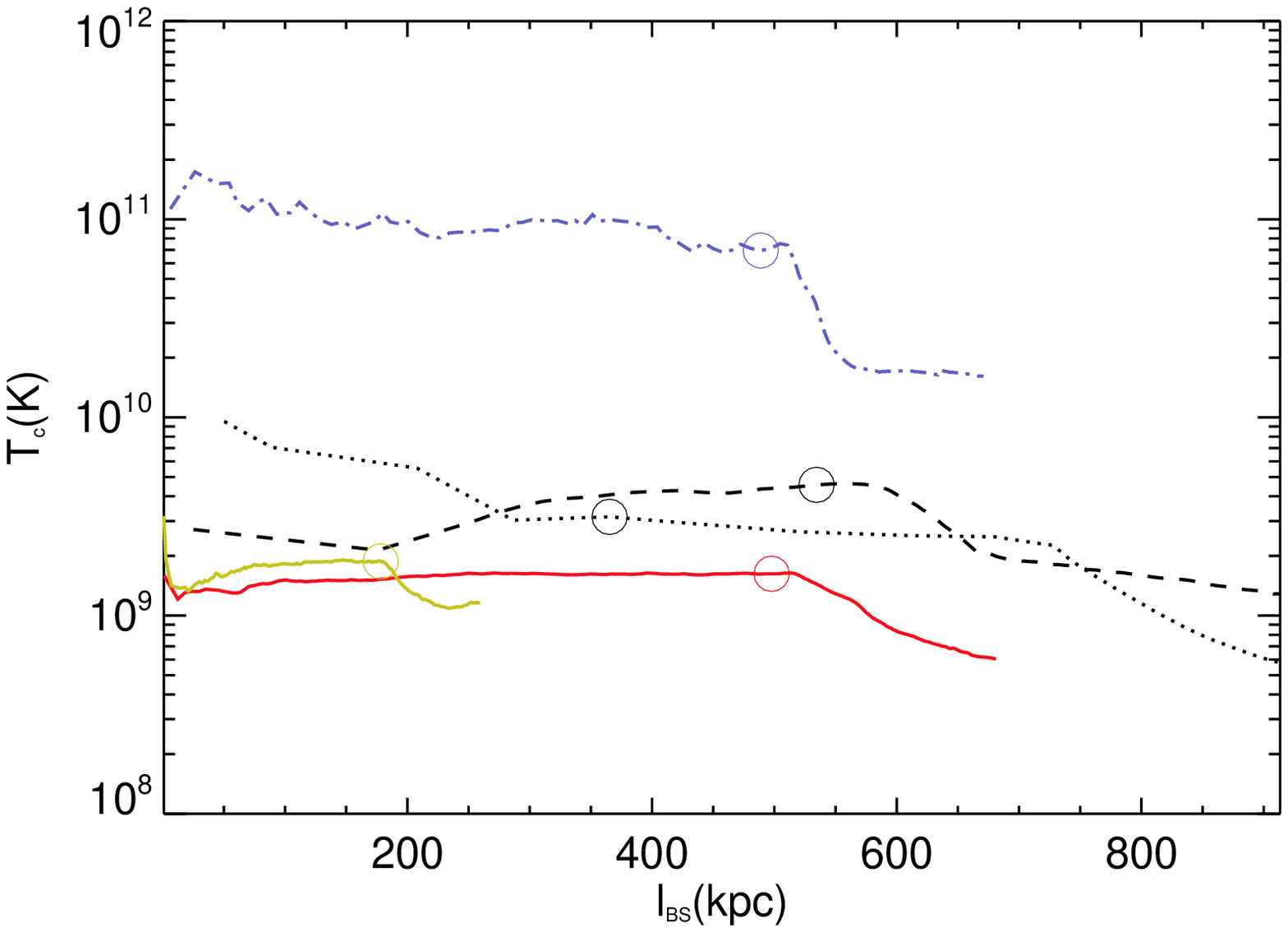}
 \includegraphics[width=0.45\textwidth]{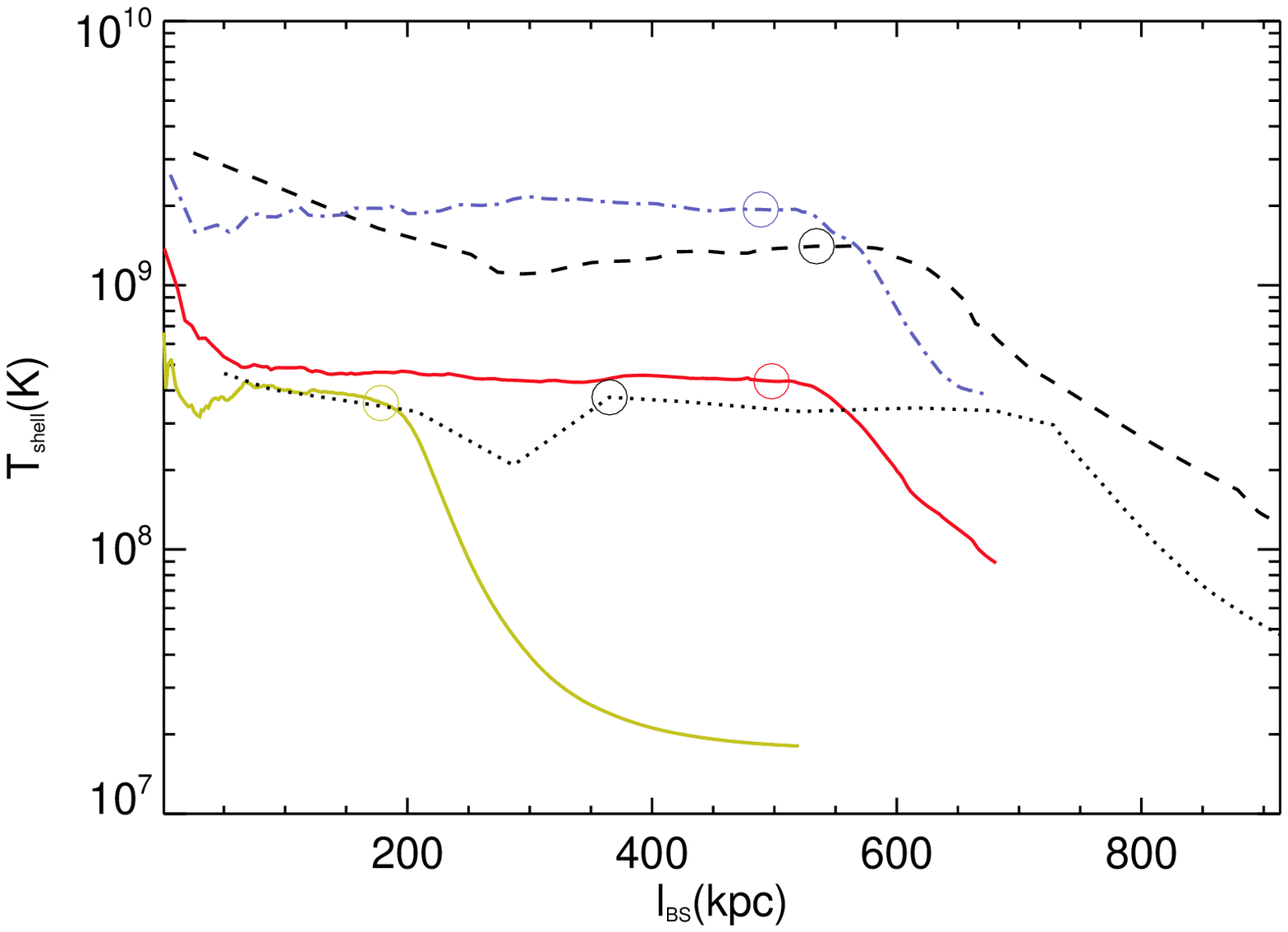}
 \caption{Cocoon (left panels) and shell (right panels) pressures (top), densities (central), and temperatures (bottom) in cgs
units, versus position of the head of the bow-shock (see
Fig.~\ref{fig:lr}). The colours indicate the different models as indicated in the top left panel. The thick, black line
in the pressure and density plots indicates the ambient medium profiles as a reference. The end of the active phase is indicated by circles.}
 \label{fig:PRT}
 \end{figure*}

Figure~\ref{fig:PRT} shows the evolution of the pressure, rest-mass density and
temperature in the shell (formed by shocked ambient medium) and in the cocoon versus the
position of the head of the bow-shock, for all the models. The cocoon is
defined as the region inside the bow-shock where the
jet-mass fraction is larger than 1\%. The shell is the rest of the
shocked volume. The drops in the evolution of the variables are caused
by the transition between the active and the passive phase (marked by
circles in the corresponding lines). During the active phase, the expansion of the
cocoon-shocked ambient medium system translates into a steep decrease of
pressure and density, whereas the temperature remains fairly constant
as expected from the analytic eBC model presented in \citet{pm07} and Paper~I. 
The pressure in the cocoon and the shell are very similar, as expected from the large sound speeds in
the shocked region, which allows a rapid homogenization of the pressure.

\begin{figure*}
 \includegraphics[width=\textwidth]{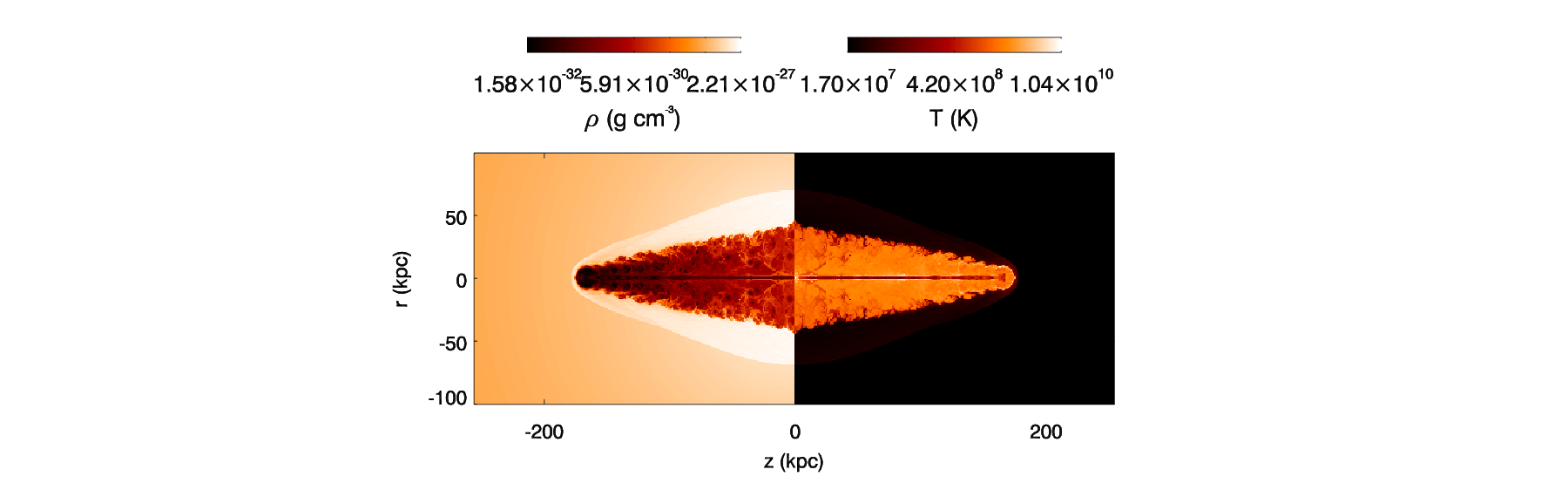}
  \includegraphics[width=\textwidth]{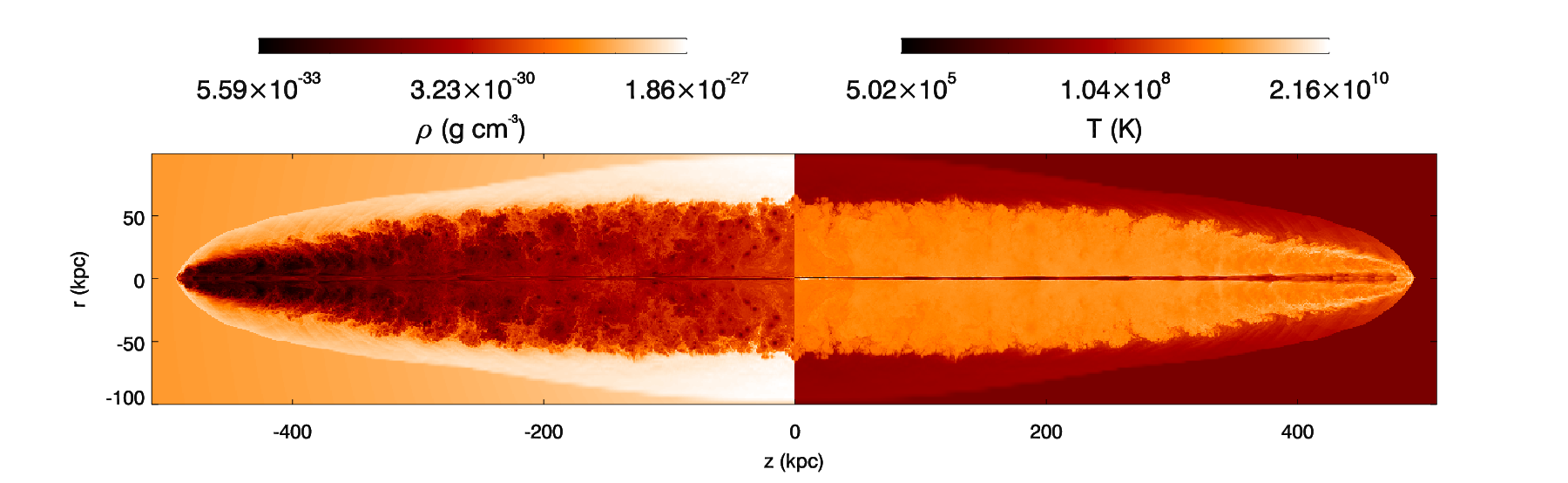}
 \includegraphics[width=\textwidth]{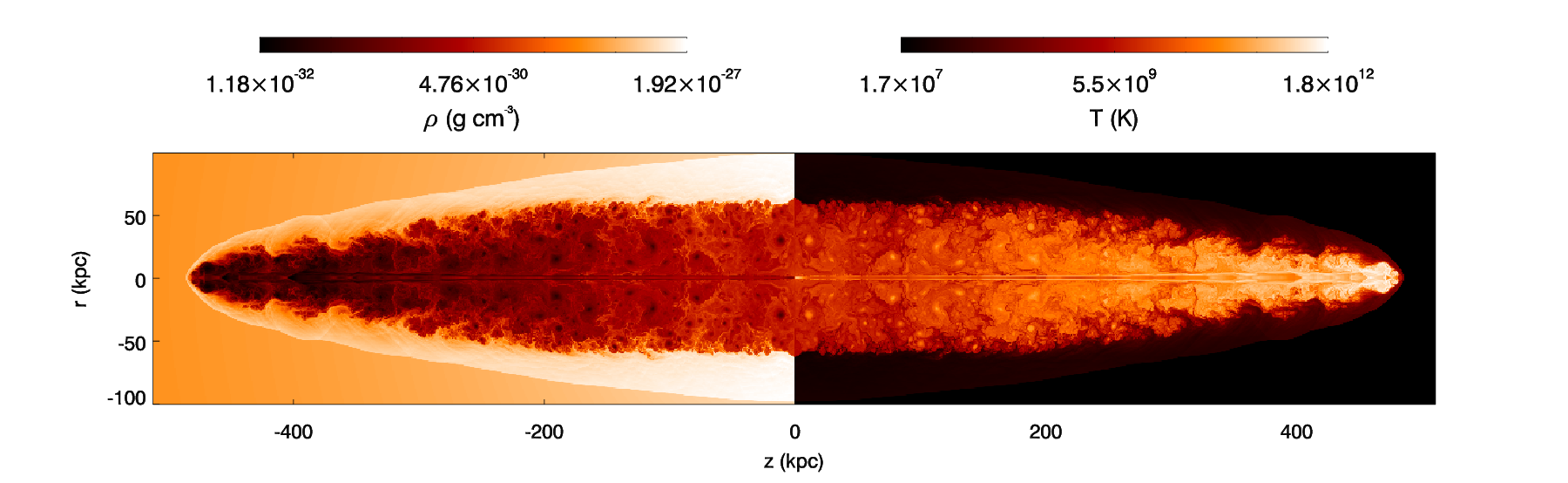}
 \includegraphics[width=\textwidth]{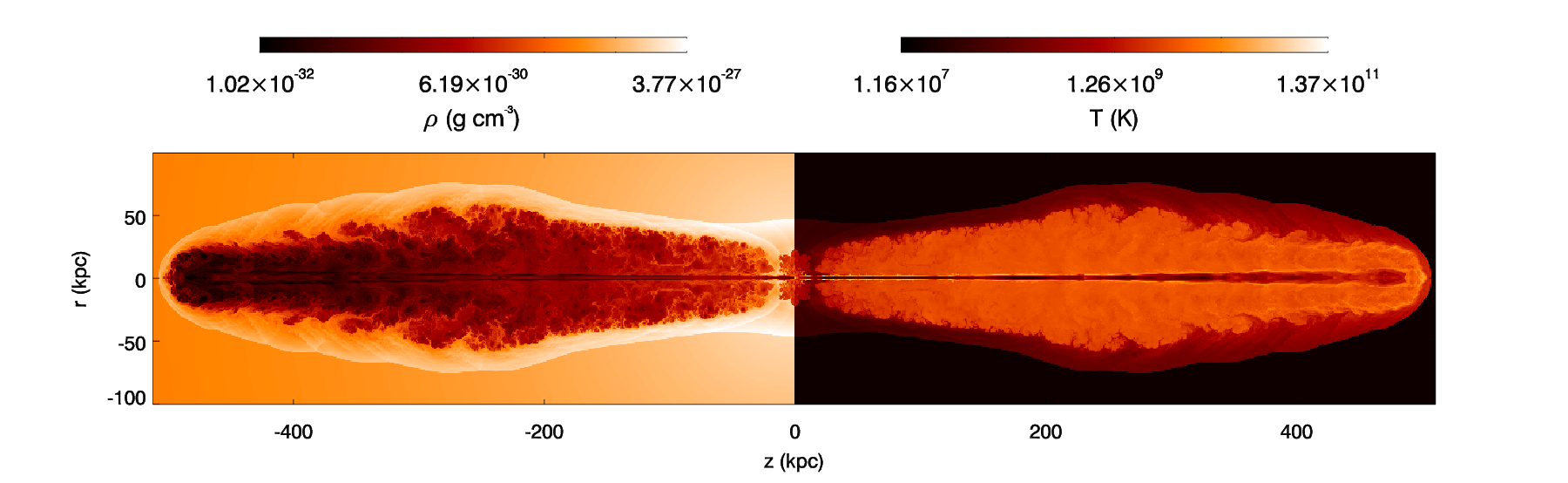}
 \caption{Density and temperature maps of models J44 (top), J45l (second row), J45b (third row) and J46
(bottom) at the end of the active phase, at $t=50$~Myr for the first three cases, and $t=16$~Myr for J46. The images have been mirrored using the two axis of
symmetry, i.e., the jet axis and the base of the jet.}
 \label{fig:astage1}
 \end{figure*}

 In the two-dimensional phase, the main differences among models are
caused by the different jet propagation speeds which control the flux of
matter and energy into the cocoon. Figure~\ref{fig:astage1} shows density and temperature maps at the
end of the active phase. These maps display the jet, the hot-spot and the
overall morphology of the bow-shock and the cocoon. Model J44 show the smaller propagation speed and thus 
has the largest relative fluxes into the cocoon and develops a wide conical cocoon and bow-shock with the smallest aspect ratio (length
over width, $l_{\rm bs}/r_{\rm bs}$). Despite its low kinetic power,
$L_{\rm k}=10^{44}\,{\rm erg/s}$, typical of powerful FRI sources, J44 shows
FRII morphology. This result follows from the suppression of three-dimensional instabilities that could disrupt the jet flow. 
The shape of the cocoon in models J45l and J45b is
similar to that of model J44 although more cylindrical towards the jet
base. The global morphology of the bow-shock and the cocoon of models
J45l and J45b (with the same jet kinetic power, density and speed but
different composition) is very similar, in agreement with the results of
\citet{sch02}. Both jets reach a distance $\simeq 500$~kpc in
50~Myr. The main difference between them lies in the values of the cocoon
temperature, which is much higher in J45b than in J45l. The densities are,
otherwise, very similar. Finally, in the case of J46, the fast initial
propagation velocity of the jet head (J46 reaches the same distance as
J45l and J45b in one third of the time) and its sudden
deceleration along the two-dimensional phase produces a slim,
cup-shaped cocoon.

%
%

\begin{figure*}
  \includegraphics[width=0.45\textwidth]{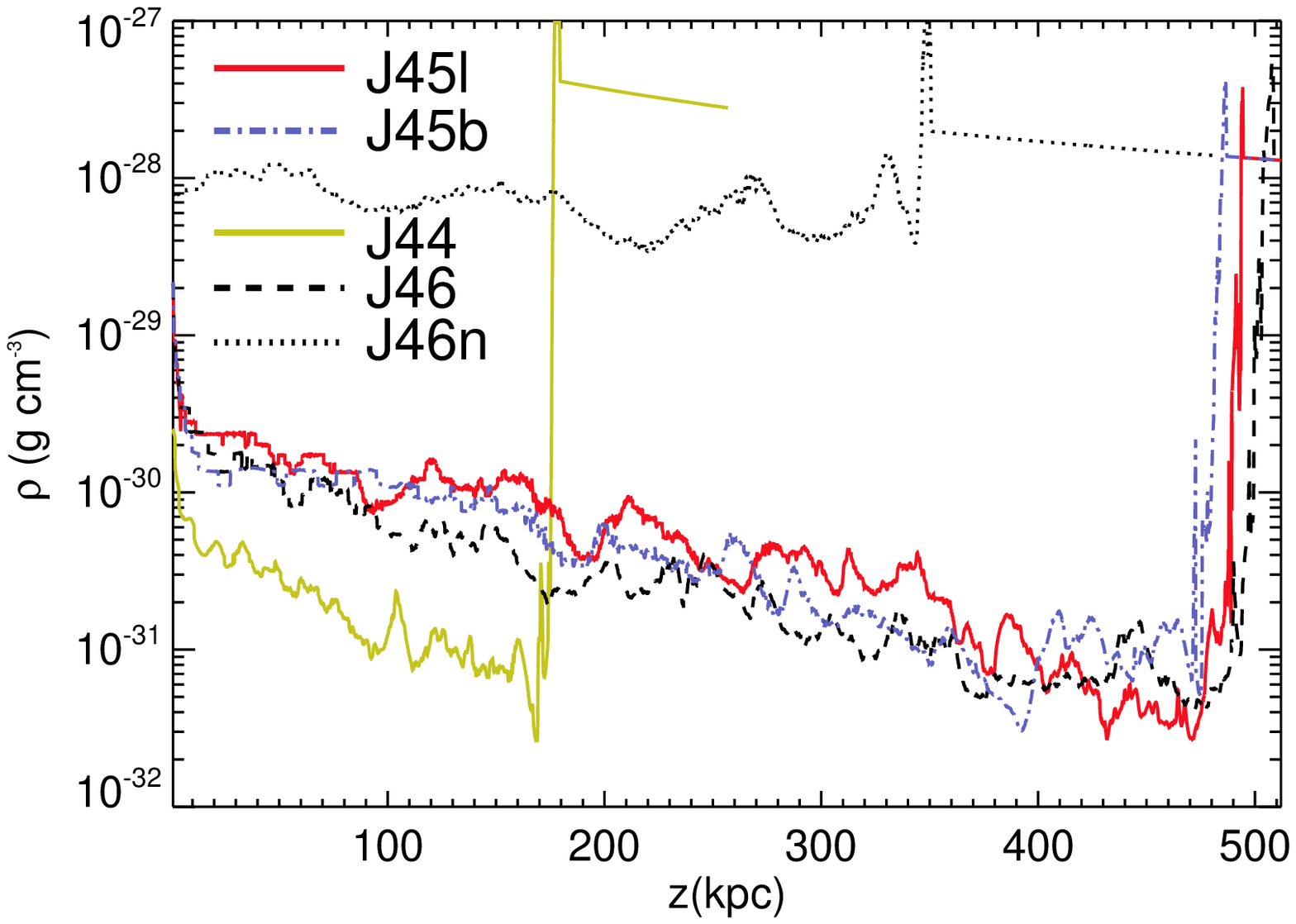}
 \includegraphics[width=0.45\textwidth]{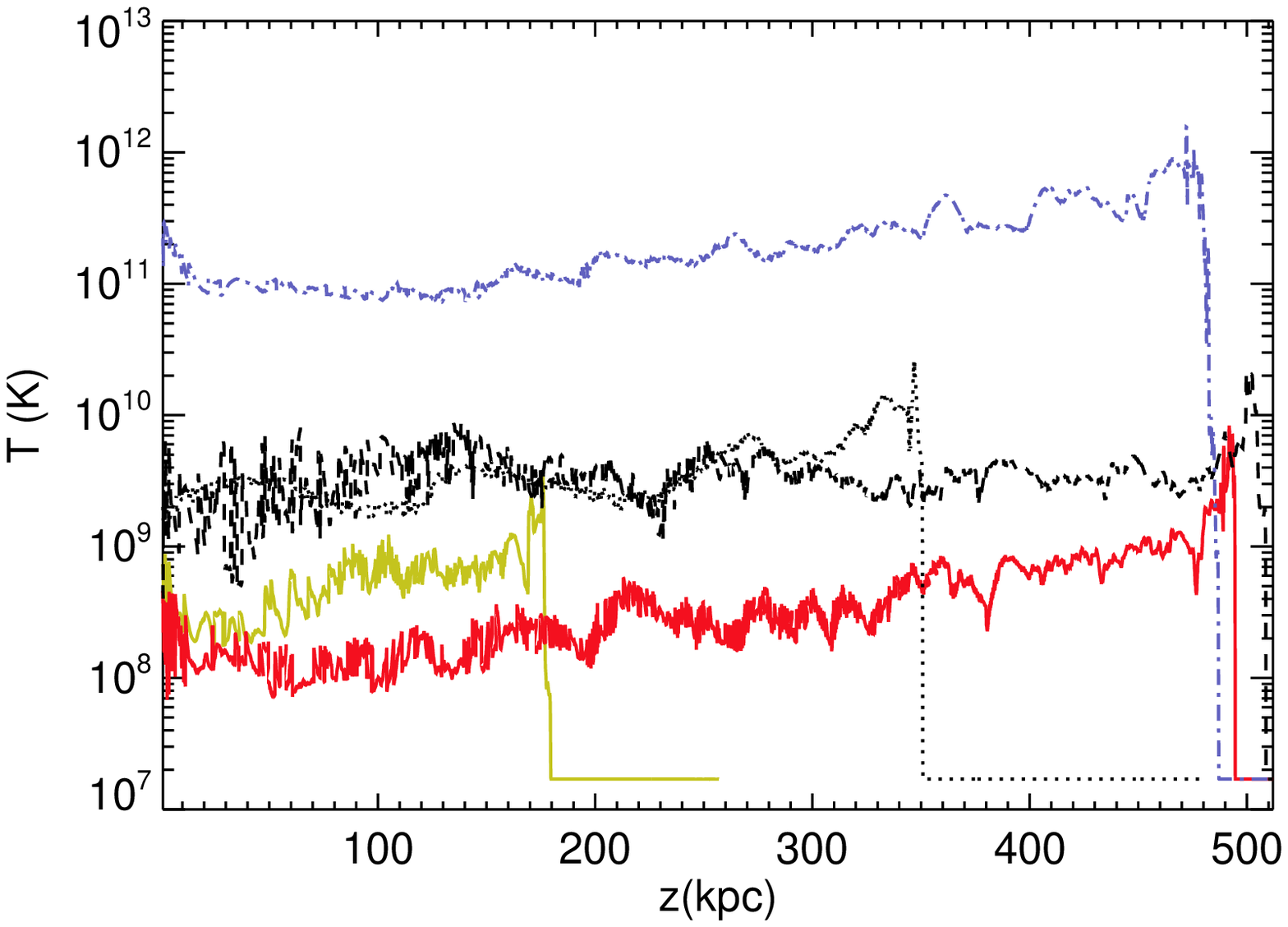}
 \includegraphics[width=0.45\textwidth]{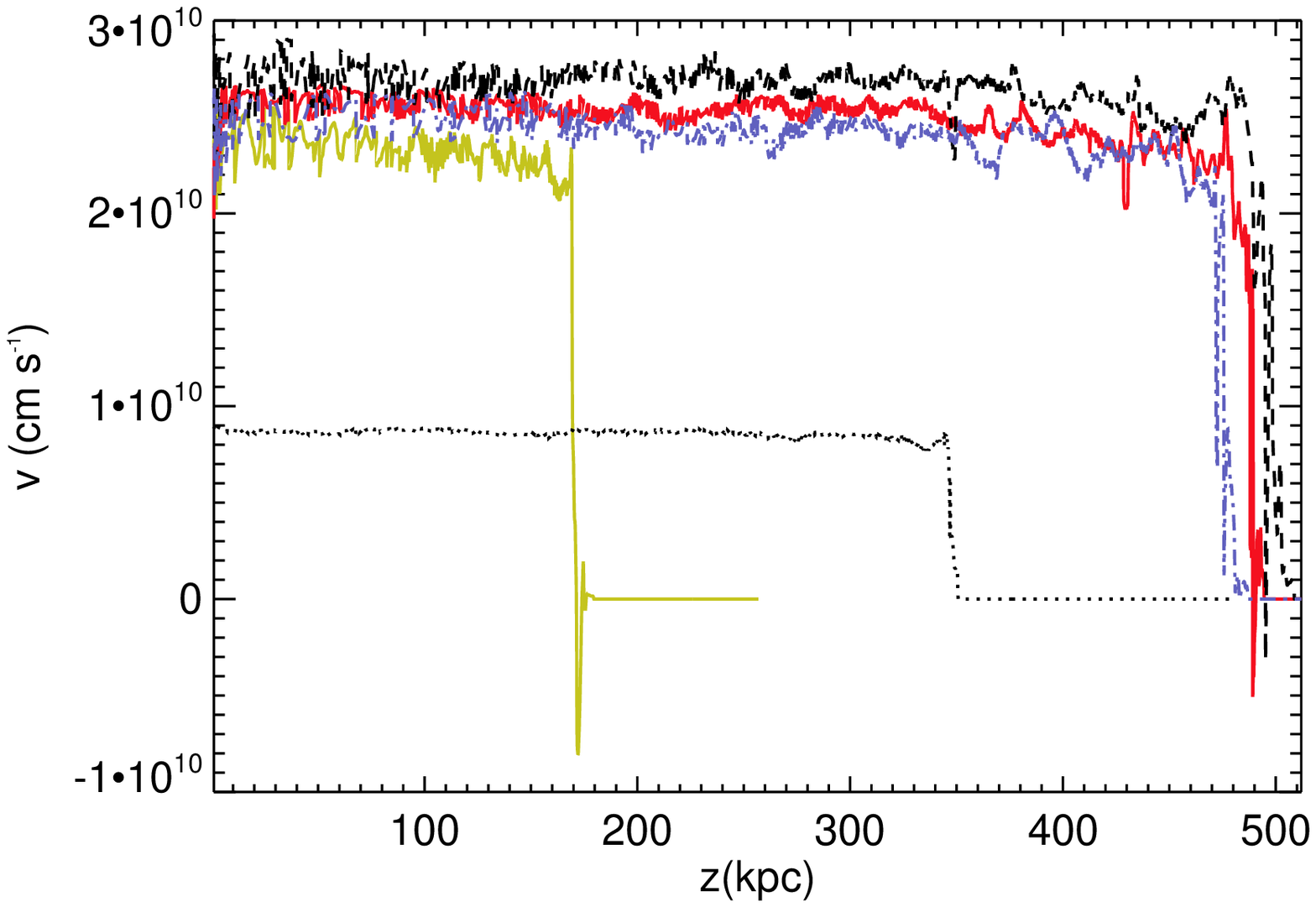}
 \includegraphics[width=0.45\textwidth]{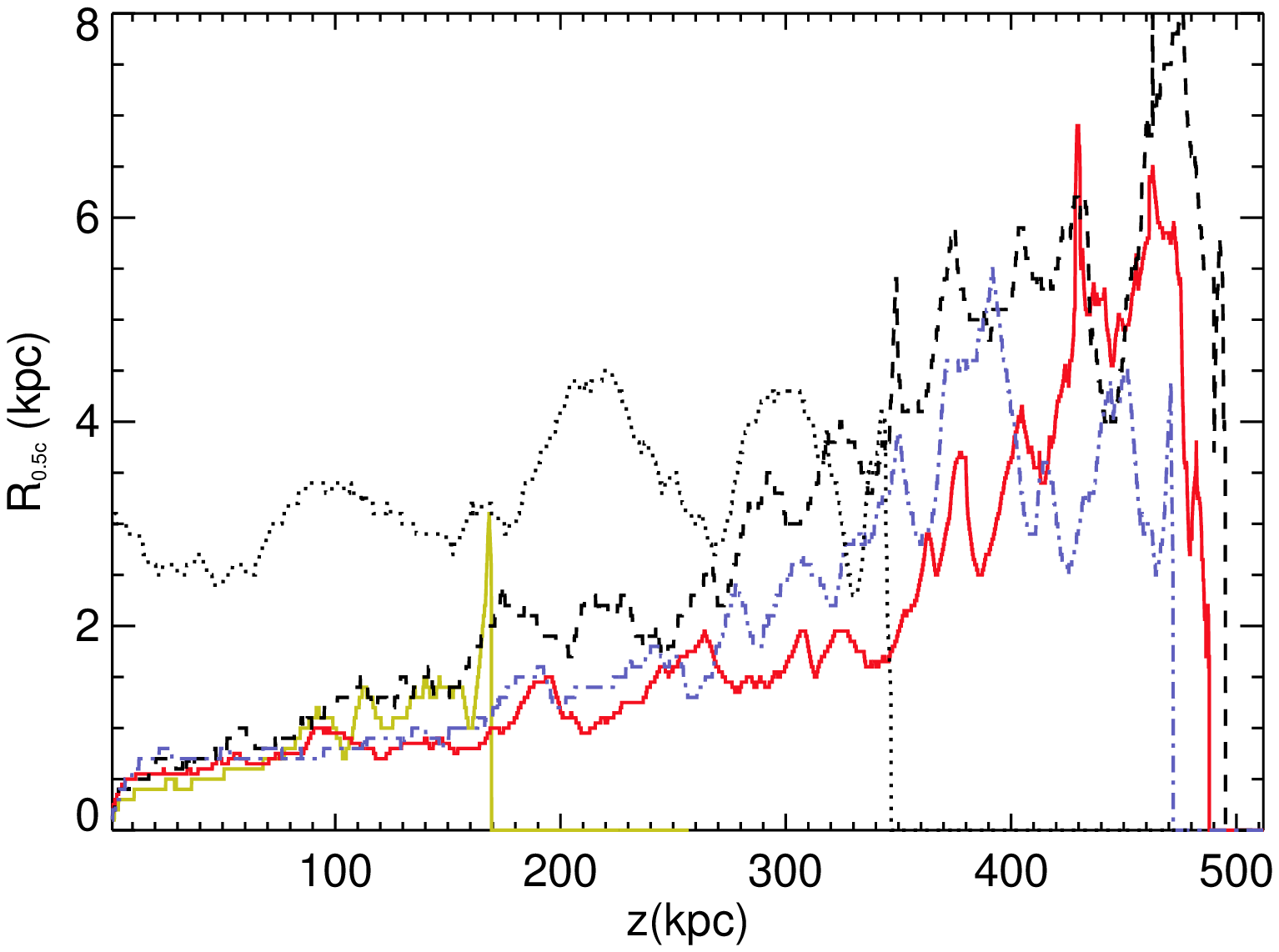}
 \caption{Mean density (top left panel), temperature (top right panel),
axial velocity (bottom left) in the jet, and jet radius (bottom right) at the end
of the active phase, coinciding with the maps shown in
Fig.~\ref{fig:astage1}. The radius of the jet
that limits the region where the mean values are calculated has been
chosen as the first position at which the axial velocity is 0.5$\,c$. Beyond the reverse-shock, the plot shows the ambient medium values of the 
parameters. The colours indicate the different
 models as indicated in the top left panel.}
 \label{fig:jprofs}
 \end{figure*}

 Figure~\ref{fig:jprofs} shows the mean values of rest-mass density,
temperature, axial velocity and jet radius versus distance. The jet
density decreases with distance (as a result of the jet widening),
with small increases at the conical shocks (reconfinement shocks due
to jet/cocoon pressure mismatch, which create a pinched jet structure
visible in the colour maps of Fig.~\ref{fig:astage1}). The temperature
rises slowly along the jet due to heating of the flow at these shocks.
The jet-flow decelerates slowly along the jet to axial velocities
$0.7\,-\,0.8\,c$ due to the entrainment of material across the
shear-layer and also to the loss of kinetic energy at the conical
shocks. However, this deceleration does not prevent the flow to reach
the reverse shock with mildly relativistic speeds. The deceleration is more
important in the case of J44, which is the less powerful model. Finally,
the jet radius grows monotonically with distance and shows the effect of
the pinching on the jet width in the continuous up-and-downs that grow
in amplitude along the jet, forced by small pressure differences with
the environment. The final jet radii at $z\simeq500\,{\rm kpc}$ is
around 5~kpc for jets J45l, J46 and J45b.

%
%

%
%

\subsection{Passive phase}

 When the energy and thrust fluxes start to decrease, the velocity of
the head of the jet and, thus, the Mach numbers of the bow shocks 
drop almost instantaneously to values between one and two (see the
bottom panel in Fig.~\ref{fig:lr}) . The reason is that the particles,
travelling at $v\simeq 0.9\,c$ along the jet need only $\simeq 
10^6$~yrs (less than $1\%$ of the simulated time) to reach the jet
head, located at 500~kpc (180~kpc in the case of J44). It is
remarkable that the Mach number of the bow-shock in J44 preserves a
value $\gtrsim 1$ over a timescale of 300~Myr.

 Figure~\ref{fig:PRT} shows that the pressure inside the shocked
region is larger by a factor $1.3 - 4$, depending on the model, than
that in the ambient medium along the whole simulation. This is the
ultimate reason for the bow-shocks not to evolve to transonic
velocities. In the case of the density, the shell and the cocoon show
very different values. The shell is formed by the shocked ambient
medium and is thus compressed and denser than the ambient medium,
whereas the cocoon is underdense with respect to the shell by a factor of a hundred to a thousand. 
The temperature in the
cocoon and the shell is fairly constant during the active phase (see
previous Section), however, once the jet has been switched-off, the
temperature in the cocoon decreases as the cocoon expands and the jet
particles share their internal energy with the cooler shocked ambient medium
particles via mixing.

 Models J45l and J45b show similar pressures and densities in the
cocoon because these parameters depend basically on the total injected
energy and mass and the volume of the cocoon, which are the same in both
cases. Regarding the values in the shell, the pressures also show a very
similar evolution. Model J45b, corresponding to the highest
temperature jet at injection gives the highest cocoon and shell
temperatures. However, the high Mach number of the bow shock in model
J46 heats the ambient medium by a large factor making the shell temperature
of this model to approach that of J45b.

\begin{figure*}
 \includegraphics[width=\textwidth]{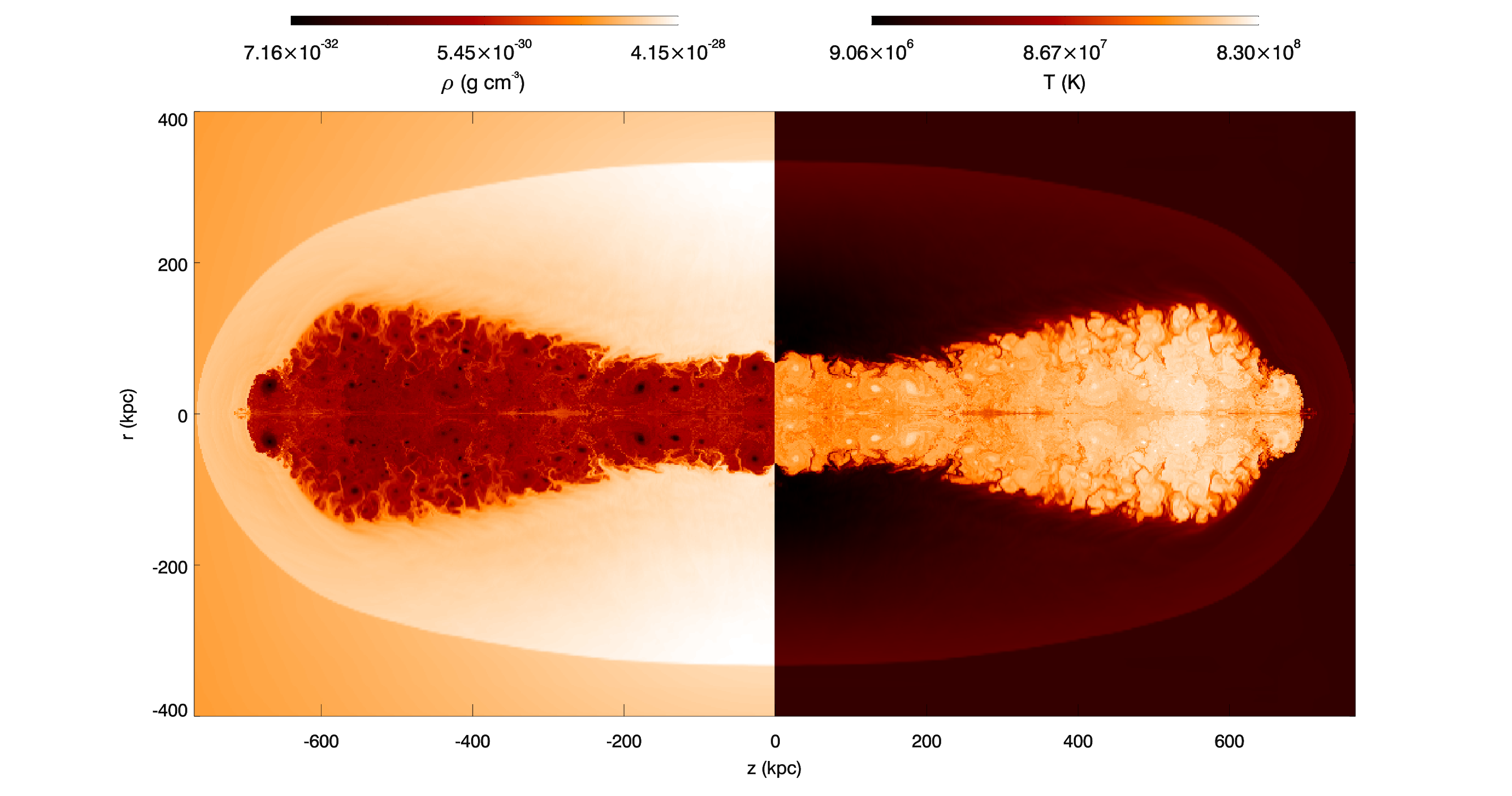}
 \includegraphics[width=\textwidth]{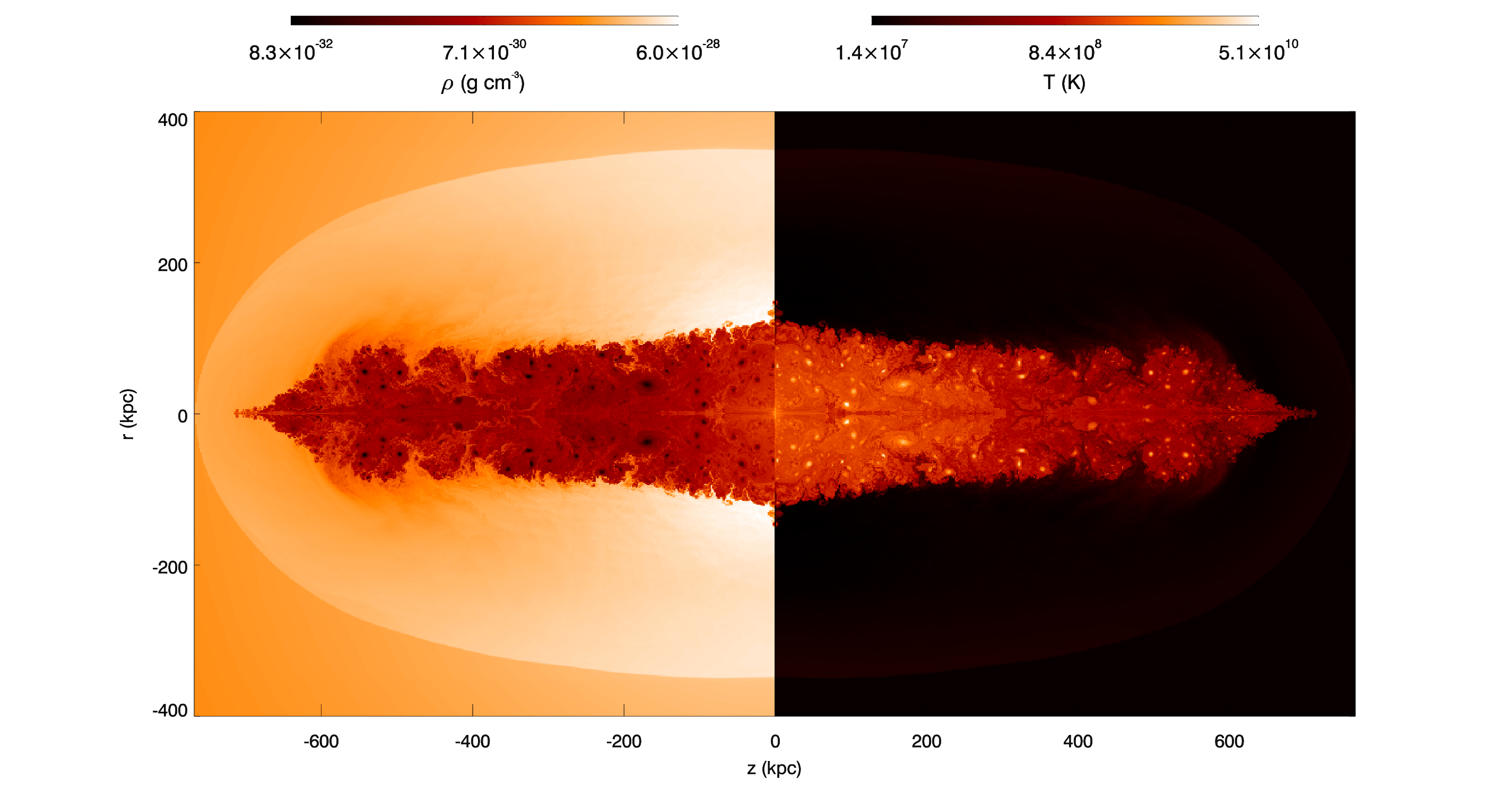}
  \caption{Density and temperature maps of models J45l and J45b towards the end of the simulations, at $t\simeq 300$~Myr and $t\simeq 250$~Myr, respectively. The images have been mirrored using the two axis of symmetry, i.e., the jet axis and the base of the jet.}
 \label{fig:pstage1}
 \end{figure*}

\begin{figure*}
  \includegraphics[width=\textwidth]{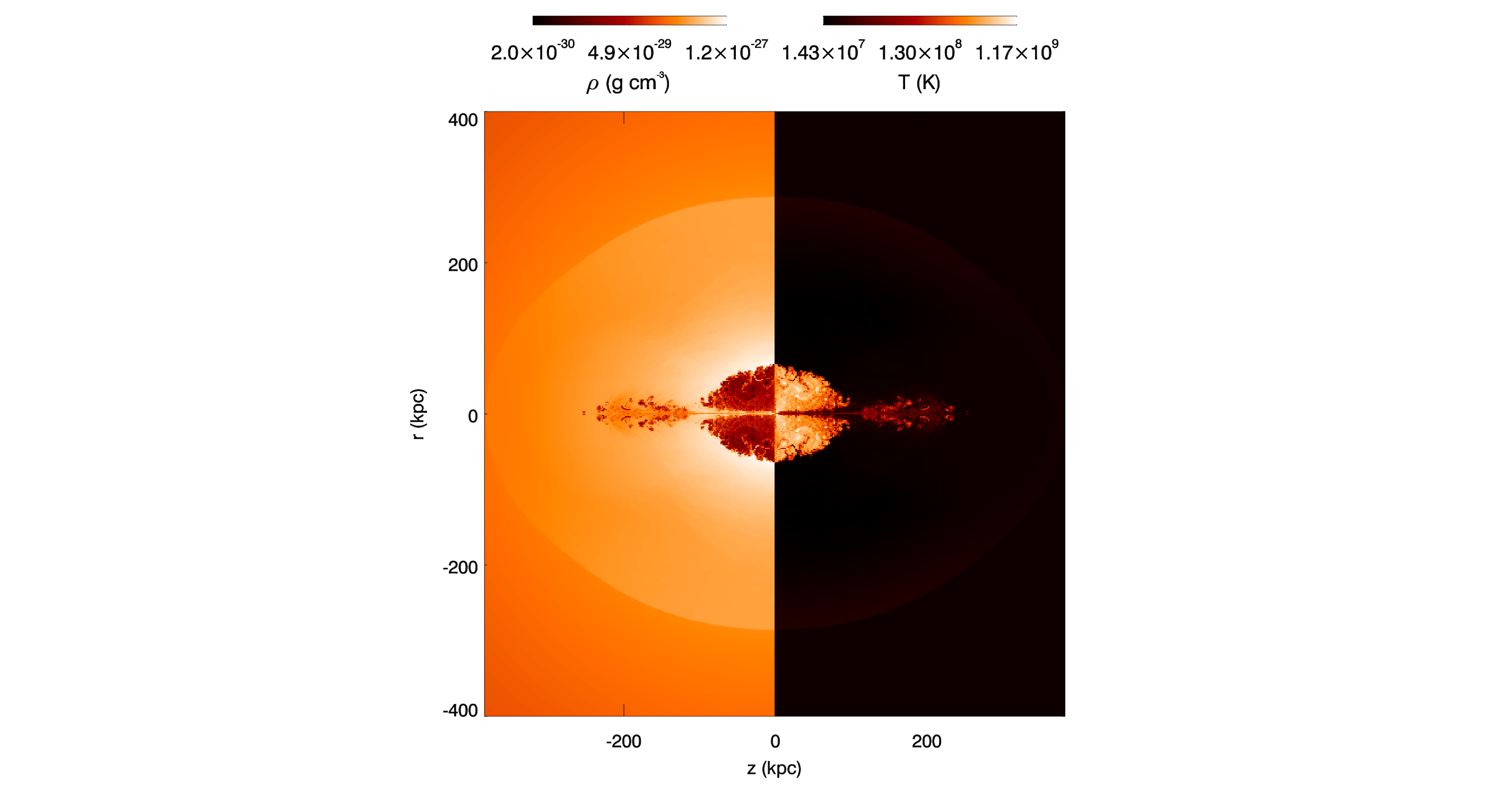}
 \includegraphics[width=\textwidth]{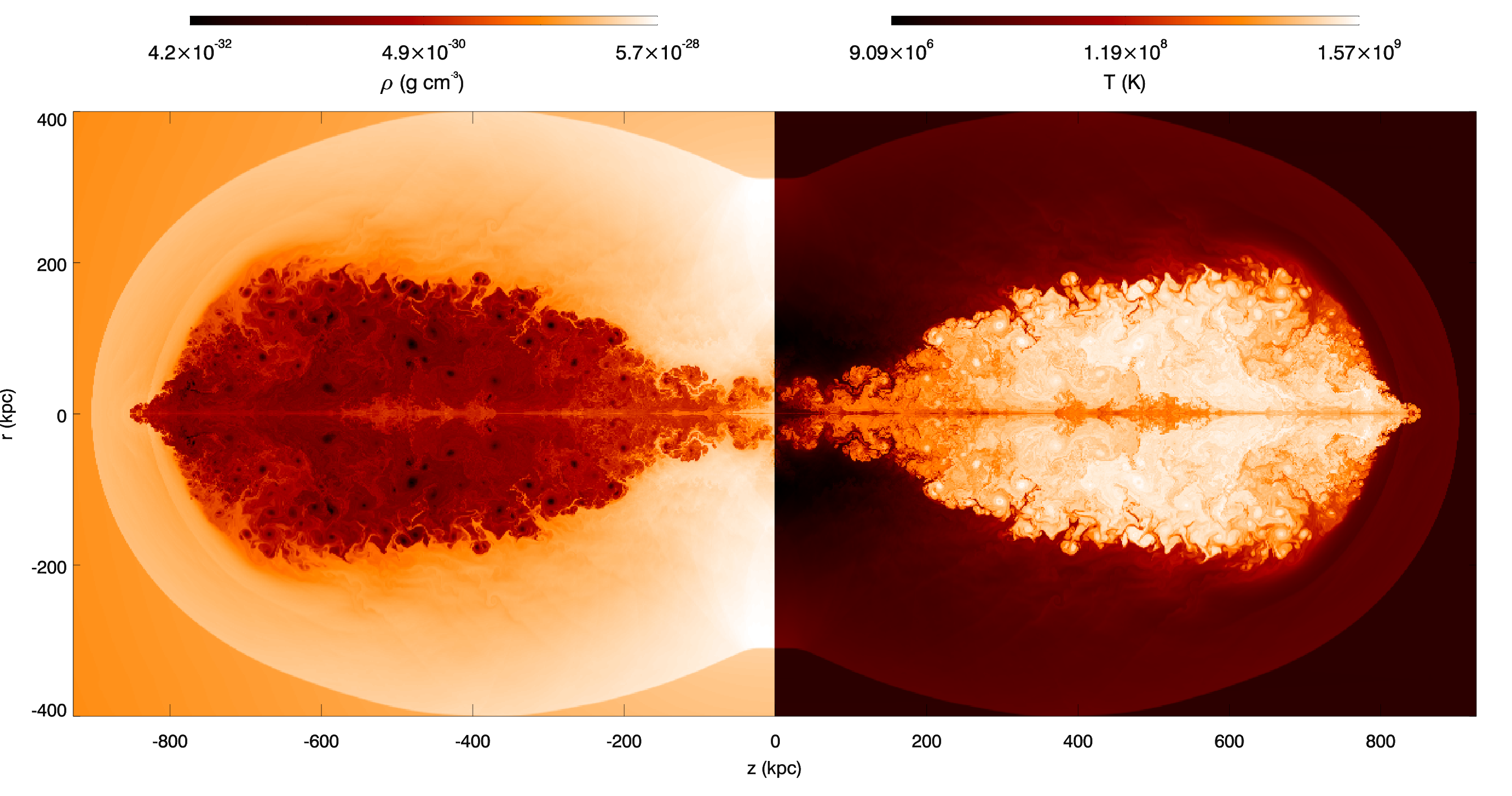}
 \caption{Density and temperature maps of models J44 (top) and J46 (bottom) towards the end of the simulations, at $t\simeq 350$~Myr and $t\simeq 180$~Myr, respectively. The images have been mirrored using the two axis of symmetry, i.e., the jet axis and the base of the jet.}
 \label{fig:pstage2}
 \end{figure*}

  Figures~\ref{fig:pstage1} and \ref{fig:pstage2} show the last snapshot
of the different simulations. The frame for J45l is at time $t=228$~Myr,
J46 at $t=184$~Myr, J44 at $t=280$~Myr, and J45b at $t=251$~Myr. The
first striking difference between J45l (leptonic jet) and J45b
(baryonic) is the different morphology of the cocoon. That of the
leptonic model widens outwards, whereas in the baryonic case it is more
homogeneous. The cocoon in J46 forms a bubble that is basically detached
from the base of the grid. On the contrary, the cocoon in J44 is divided
into two parts. The main one is formed by matter deflected at the
reverse shock at early times in the simulation. The slow advance speed
of the jet's head and its low power places, via backflow, large amounts
of matter close to the center which are isotropized by the gravitational
field. A similar effect is seen in other simulations of low power jets
in the context of models of deceleration of FRIs \citep{pmlh14}. The second piece of the cocoon lies around the old jet
axis and is formed by the last ejected particles, which do not
flow back after passing through the reverse shock because this shock has already disappeared or becomes very weak as the injection flux decreases. 

%
 %

%
 %
 
\subsection{Non-relativistic outflow}

%
\begin{figure*}
 \includegraphics[width=\textwidth]{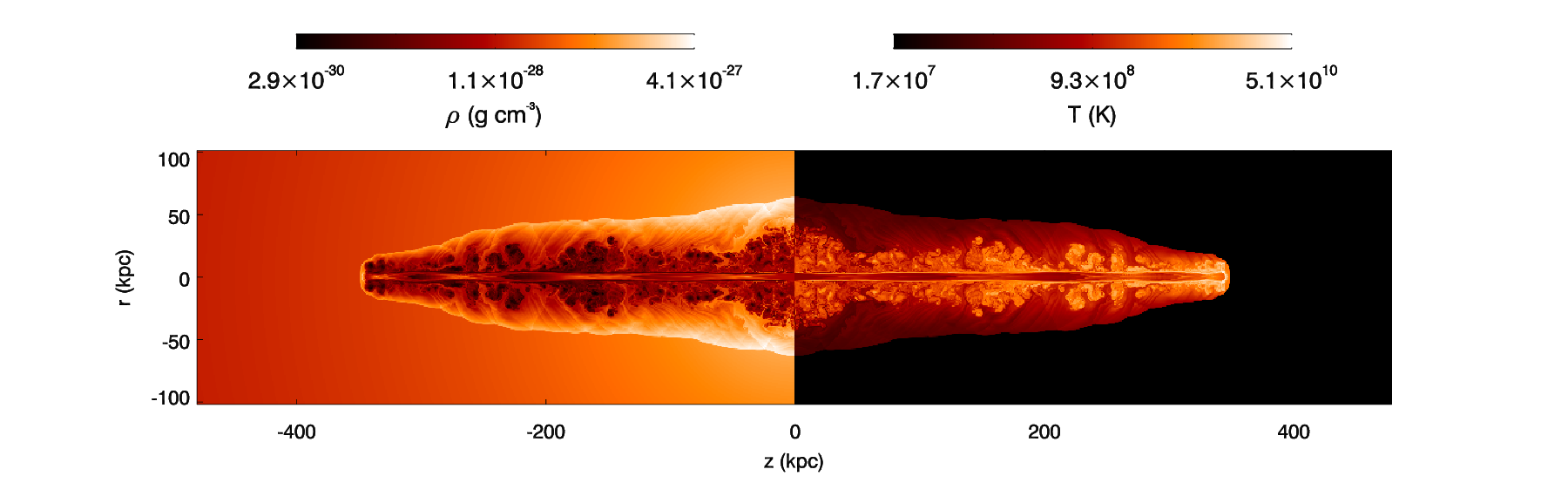}
 \includegraphics[width=\textwidth]{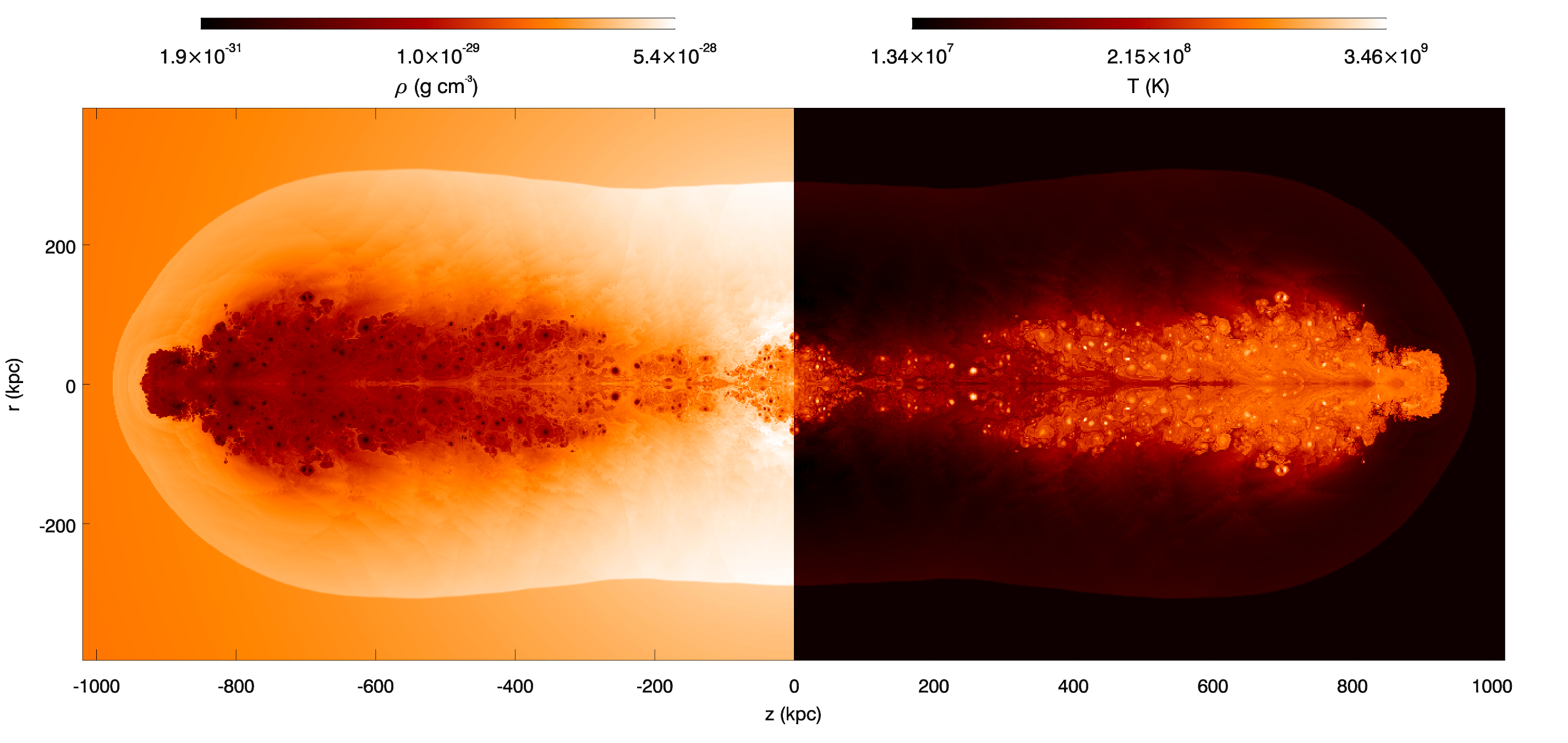}
 \caption{Density and temperature maps of model J46n at the end of the active phase, $t=16$~Myr (top panel), and at the end of the simulation, $t\simeq170\,{\rm Myr}$ (bottom panel). The images have been mirrored using the two axis of symmetry, i.e., the jet axis and the base of the jet.}
 \label{fig:astagen}
 \end{figure*}
%


    The black, dotted lines in Fig.~\ref{fig:PRT} indicate the shell and cocoon pressure, density and
temperature for the slow jet model J46n. As seen in the corresponding panels, the cocoon and shell pressures
in models J46 and J46n differ by a small factor. This is expected because the
pressure is basically a result of the total energy input (ignoring the
rest mass energy) divided by the total volume. The volume of the
cocoon/shell system is smaller in model J46n, but the internal energy
is larger in the case of J46. The shell densities of both models
are also similar since the outer shock is strong and the classical limit
in the density jump across the shock is reached in both cases. However
the cocoon density is larger in the slow jet because the total mass flux
is much larger in this case. Taking into account that the jet density is
the same and the jet radius is a factor 30 larger in this case, the mass
flux is 900 times larger in the slow jet than in J46. Finally, the
bottom panels of Fig.~\ref{fig:PRT} show that the cocoon presents a similar 
temperature in both cases and that the shell temperature is larger in J46. Regarding the similar temperature in the cocoon,
it can be explained as in the case of models J45l and J45b. Since
the slow jet is made of electrons and protons, the number of injected particles
is smaller and the energy per particle is increased, even though the slow jet is colder. As a result, the temperature of the
slow jet is 30 times larger than that of J46 at injection. Once in the cocoon, the temperature
changes and the difference between the two models is
reduced due to mixing with the ambient medium particles. Consistently with the fact that the pressure is slightly larger
and the density slightly smaller in the relativistic model, the energy per
particle and hence the temperature is few times larger in the shell of
model J46. A clear difference between both models is the time lapse
between the drop in jet injection (indicated by circles in the
panels) and the start of the corresponding decrease in density and
pressure. This is due to the lower flow velocity in the non-relativistic
jet. The last injected particles take more time to reach the jet head
(where they get into the cocoon) in the case of J46n. The different nature 
of the flow with respect to the relativistic jets is clearly seen in Fig.~\ref{fig:jprofs}. The non-relativistic flow
is denser, slower and shows less structure, as expected from its larger inertia. This structure does
not recover, by construction, the observed properties of powerful FRII jets, which are mildly relativistic up to hundreds of
kiloparsecs, as shown by the brightness asymmetries between jets and counter-jets in those powerful jets \citep[see, e.g.,][]{bri94}.

 Figure~\ref{fig:astagen} shows the maps of rest-mass density and temperature for model J46n at the end of the active phase ($t\simeq16\,{\rm Myr}$, top panel) and at the end of the simulation ($t\simeq170\,{\rm Myr}$, bottom panel). The non-relativistic jet is slower than its relativistic counterpart during the active phase as deduced from the positions of the jet head of
J46 (bottom panel of Fig.~\ref{fig:astage1}) and J46n (top panel of Fig.~\ref{fig:astagen}). Conversely, during the passive phase, the
axial expansion of the cocoon/shell system becomes faster (and the
sideways expansion slower) in J46n owing to the larger inertia of the
protons that forms the non-relativistic jet (see bottom panels of 
Figs.~\ref{fig:pstage2} -J46- and \ref{fig:astagen} -J46n-). Thus, an important difference between
leptonic dominated and baryonic dominated jets is the aspect ratio of
both the outer shock and the cocoon, which is smaller in the relativistic case.

\section{Discussion}
\label{disc}

\subsection{Cocoon evolution}

 In Paper~I, the long-term evolution of the cocoons in our numerical
simulations was interpreted within the so-called extended
Begelman-Cioffi's model \citep[eBC from now on,][]{bc89,sch02,pm07}, which
describes the expansion against the ambient medium of the
overpressured cocoons raised by the continuous injection of energy
from a supersonic jet. In this model, the axial expansion of the
cocoon (i.e. along the jet) proceeds at the advance speed determined
by the jet, whereas the sideways growth follows from the assumption of
the evolution being mediated by a strong shock. The model allows for a 
power-law dependence of the jet advance speed with time and a
non-uniform ambient medium described by some power law. In 
addition, the model can also describe the passive (supersonic)
expansion of the cocoon once the jet has ceased its activity (Sedov
phase).

 As concluded in that paper, the eBC model describes consistently the
long-term evolution of the simulated cocoons along the jet active
phases (phases i) and ii); see Sect.~\ref{s:ap}). In the present paper, we concentrate in the 
Sedov phase, where the differences between the model and the simulations are larger.
In this late phase of the evolution, the sideways expansion seems to
be better described as being mediated by a weak shock for which the
expansion speed is$\approx c_{s,{\rm c}} \propto (P_{\rm c}/\rho_{\rm
  c})^{1/2}$ (where $c_{s,{\rm c}}$, $P_{\rm c}$ and
$\rho_{{\rm c}}$ are the mean sound speed, pressure and  
rest-mass density of the cocoon, respectively). In the limit of adiabatic expansion,
$c_{s,{\rm c}} \propto P_{\rm c}^{1/5}$ (for an adiabatic exponent of
5/3), which introduced in the eBC model gives $R_{\rm c} \propto t^{(4-\alpha)/7}$ (where $\alpha$
is the power of the advance speed of the bow shock along the axial
direction). Computing the exponents of the sideways expansion with time for the 
models listed in Table~2 of Paper~I, one gets approximately $R_{\rm c}
\propto t^{2/3}$ for the three cases ($0.66$-$0.69$), still smaller but closer to the
simulated values ($0.72$-$1.00$) than the ones obtained assuming a strong shock 
wave in the sideways expansion ($0.54$-$0.61$). 

 We now discuss the role of buoyancy as the driving mechanism for
the passive cocoon expansion. The force of buoyancy experienced by a
bubble of plasma in a gravity field, for large ambient medium to bubble
density contrast, is proportional to the volume of the bubble, the ambient medium
density and the strength of the local gravitational field. We
can estimate the role of buoyancy in the expansion of the underdense
cavities\footnote{The underdense region defined by the cocoon
  (Fig.~\ref{fig:PRT}) corresponds to the X-ray cavity related to the
  regions filled by the radio-lobes formed by jets (see, e.g.,
  Paper~I).} formed by the jets by describing those cavities as pairs of bubbles
(the cocoons of the jet and the counter-jet) separating
from the center of the gravitational potential well. The basic
ingredients of the model are the distance of the bubbles to the
galactic center, $L_{\rm b}$, the radius of the bubble, $R_{\rm b}$,
the ambient medium density, $\rho_{\rm a}$, and the acceleration of
gravity. The rise velocity of the bubbles,
$v_{\rm b}$, is then
\begin{equation}
\displaystyle{v_{\rm b} \propto t^{ - \frac{3 \delta +2}{\beta + \gamma}
- 1}}.
\label{buoyancy:v_b}
\end{equation}
In this simple model, the volume of the bubble is proportional to
$R_{\rm b}^3$, with $R_{\rm b} \propto t^\delta$, whereas the
gravitational acceleration is the one created by the dark matter halo,
with a spherical density distribution given by $\rho_{\rm DM} \propto
r^\gamma$. For the ambient medium density we have $\rho_{\rm a} \propto
r^\beta$. In our numerical setup $\beta = -1.02$ beyond 100 kpc. 
On the other hand, $\delta$ varies typically between $2/5$ for a
strong shock (i.e., Sedov expansion) and $1/2$ for a weak shock/transonic 
expansion (assuming adiabatic expansion, with adiabatic index 5/3). 
Finally, for a NFW profile of dark matter distribution, $\gamma 
\approx -1$ for $r$ much smaller than $R_{\rm s}$, the halo {\it scale radius}, 
$-2$ for $r \approx R_{\rm s}$, and $-3$ for $r \gg R_{\rm s}$. Beyond 100~kpc, the 
dark matter profile used in our simulations can be fairly fitted by a NFW
profile with $R_{\rm s} \approx 270$ kpc and $\gamma$ can be taken 
$\approx -2$ up to 1 Mpc \citep[see also][]{ko99}.

Equation (\ref{buoyancy:v_b}) leads to accelerating bubbles for almost
all the combinations of parameters within the ranges given above. In particular,
taking $\{\beta,\gamma,\delta\} = \{-1, -2, 1/2\}$, the rise velocity of
the bubbles follows $v_{\rm b} \propto t^{1/6}$ (i.e., slightly accelerating), which is
in conflict with the law for the axial expansion velocity of
the cavity ($t^\alpha$ with $\alpha = -0.83, -0.60$; see Table~2 of
Paper~I) found in our simulations. The conclusion is that buoyancy is
not the dominant process causing the expansion of the cavity.

The model described in the previous paragraphs does not take into
account the effect of dragging onto the bubble, which in the long term
will cause it to reach some limit speed. Hence, another way to
look for clues of buoyancy is to estimate the limit speeds and to
compare them with the actual cavity expansion speeds. The terminal speed can 
be obtained by balancing the 
buoyant force and the drag \citep[see, e.g.,][]{ch01}
\begin{equation}
v_{\rm b, t} \approx \sqrt{2g\frac{V_{\rm b}}{S_{\rm b}}},
\end{equation}
where $V_{\rm b}$ and $S_{\rm b}$ are the volume and the
cross section of the bubble, respectively, and $g$ is the acceleration of gravity,
all assumed to be constant. In this expression, the drag coefficient, 
depending on the geometry of the bubble and the Reynolds number, is
probably overestimated by setting it to 1. Now, applying this
expression to model J46 close to the jet switch-off ($t = 13.5$ Myrs) and
at the end of the simulation ($t = 180$ Myrs), we get the terminal
speeds $v_{\rm J46, t} = 3.75 \times 10^{-3} c$, $3.15 \times 10^{-3} c$,
where the variation comes from the changing conditions along the bubble
evolution. On the other hand, at $t = 13.5$ Myrs, the measured rise speed of 
the bubble is larger than $6.4 \times 10^{-3} c$. The fact that
the true rise speed is about a factor of 2 larger than the
expected terminal speed by drag should be interpreted as an additional
sign that another process different from buoyancy (in this case the
Sedov expansion) is the driver of the cavity evolution. However, the same exercise 
repeated for the less powerful jet model J45l gives an average rising speed of 
$1.36 \times 10^{-3} c$, well below the estimated limit speed, 
$\approx 4.30 \times 10^{-3} c$. Therefore, the previous argumentation cannot be applied 
in this case. 

  A possible picture emerging from the results presented in this section is that 
the expansion of the cavities in our simulations is still driven by shocks in all 
the cases, but in contrast with earlier phases, the evolution of the cavities along 
the passive phase is better described as mediated by weak shocks. This 
effect is more apparent in the less powerful jets, which can be undergoing a gentle 
transition to buoyancy.

\subsection{Large-scale morphology: comparison with observations}
%
\begin{figure}
 \includegraphics[width=0.45\textwidth]{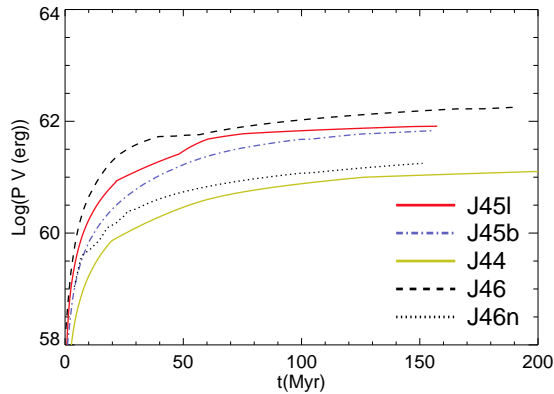}
 \caption{Work done by the injected gas to displace the ambient medium, considering the whole volume of the cocoon. The colours of the lines indicate the different models as indicated.} 
 \label{fig:we}
 \end{figure}
%
\subsubsection{Jet composition and thermodynamics}

Figure~\ref{fig:we} shows the work done to create the cavities for all models. In the case of the relativistic jets, this represents a large amount of the injected energy. There is, however, a difference in the work performed in terms of the original jet composition (dash-dotted blue and solid red lines in Fig.~\ref{fig:we}). The reason for this difference is the larger relative inertia of the protons that dominate the jet dynamics in J45b, which results in a larger axial expansion, as opposed to the important sideways expansion in the case of J45l. This is also an important difference between models J46 and J46n. J46n shows that the inflation of a cavity is necessarily related to the value of the internal energy in the jet with respect to its kinetic energy. 

  A second major difference between the relativistic and the non-relativistic jets (J46 and J46n) concerns the amount of gas showing an increase in entropy. The entropy follows accurately the temperature maps (Figs.~\ref{fig:astage1}, \ref{fig:pstage2}, and \ref{fig:astagen}) and, although the average value of entropy is similar in both cases, the volume of gas with increased entropy is much larger in the relativistic case.

 \subsubsection{Cocoon composition and overpressure}
 
  Cocoons are composed by a mixture of shocked jet gas and shocked ambient gas. The mixing occurs via Kelvin-Helmholtz instabilities growing at the contact discontinuity between both shocked flows. The mass contribution of the shocked ambient medium is larger, even though only a small fraction of the total amount of shocked ambient medium is entrained in the cocoon. Despite the contribution of shocked ambient gas, the cocoon is hotter than the shocked ambient medium, owing to the large temperature of the jet particles, which are not only injected with high temperature but also gain internal energy at the recollimation shocks along the jet and at the reverse shock, during the active phase. A number of works have studied the composition of cocoons in FRII jets. \citet{cro04} showed that the energetics in the lobes of FRII jets require the existence of a cold gas component or magnetic field that dominates the pressure if it has to be overpressured with respect to the ambient medium. If we consider that the radio and X-ray emitting particles have to be at the high-energy tail of the particle distribution, these emitting particles can only be shocked jet particles that have been accelerated at the reverse shock or within the turbulent cocoon. However, there may exist a non-emitting population of thermal particles. \cite{ito08} have suggested that the thermal particles could dominate the cocoon pressure. Their result indicates that the internal energy budget of the non-thermal particles is larger than that of the magnetic field by a factor of a few, whereas the total internal energy of the thermal component could be up to two orders of magnitude larger. \cite{co08,co11,co13} have shown that the contribution of this thermal component to the cocoon energetics could be determined by measuring the Sunyaev-Zeldovich effect in the radio lobes.
    
   We observe that the contribution of the jet flow particles to the total amount of particles in a cell within the cocoon is smaller than 10\% at any time, which implies that, even considering that all jet particles were non-thermal, there would still be a factor 10 more particles mixed from the shocked ambient medium. Therefore, we conclude from our simulations that mixing with the shocked ambient gas is efficient enough to compensate for the apparent low pressure in the lobes measured from the non-thermal population. 
  
  This conclusion is relevant from the perspective of heating, too: as the cocoon (cavity) pressure is measured from the non-thermal component alone, the broadly spread idea that cavities are in pressure equilibrium with the ambient medium could be inaccurate. This possible misconception is used in the literature to exclude the possibility of weak shocks surrounding X-ray cavities and to interpret outward cavity expansion in terms of buoyant motion, a regime that has not been reached by our jets after very long simulated times. Although a strongly magnetized cocoon seems to be excluded by different authors \citep[][and references therein]{ito08,co11,iso11}, it is important to remark the effect that it could have on the picture drawn in the previous paragraphs. A dynamically important field could prevent mixing at the contact discontinuity between the shocked ambient medium and the shocked jet gas (if aligned with this discontinuity), thus reducing the contribution of this thermal gas to the cocoon/cavity pressure. Taking into account that the derivation of the non-thermal component of pressure from observations takes the intensity of the magnetic field into account, this would bring the estimates of the cavity pressures based on the non-thermal component alone closer to the real values of lobe-pressure, thus pointing to pressure equilibrium between cavities and ambient medium and, consequently, the presence of shocks could be ruled out.

  \subsubsection{Comparison with observations}
      
       Our simulations should be compared with those sources in which weak shocks have been detected, which typically correspond to high-power jets, one of them showing FRII morphology \citep[e.g., 3C444,][]{cro11}. These shocks have been shown to propagate to kiloparsec scales in powerful radio sources, like Hercules~A \citep{nu05}, Hydra~A \citep{si09b}, MS0735.6+7421 \citep{mc05}, HCG~62 \citep{git10}, 3C~444 \citep{cro11} or PKS B1358-113 \citep{sta14}. Our simulations reproduce the existence of large-scale, low Mach number shocks around powerful radio sources. Simulation J46 recovers their large-scale morphology \citep[e.g.,][and Paper I]{mc05,cro11}, with the outer shock pinched at the center. The similarity is improved towards the end of simulations, when J46 is long non-active, although in the case of the observed sources they appear to be active \citep[at least in the case of 3C~444,][]{cro11}. We expect that J46 would still recover the observed structure during the active phase if 1) the jet injection power was reduced, 2) the ambient medium density was larger or 3) three-dimensional instabilities developed (that are forbidden in the case of our axisymmetric simulations), as one of the causes that results in that morphology is a decrease in the head advance velocity. However, the deceleration of the jet head should not be enough to change the global morphology to that obtained in the case of J45l, J44, J45b and J46n. We thus expect that fine-tuning of the jet and ambient medium properties will result in closer simulated evolutions to those observed. 
       
\begin{figure*}
\includegraphics[width=\textwidth]{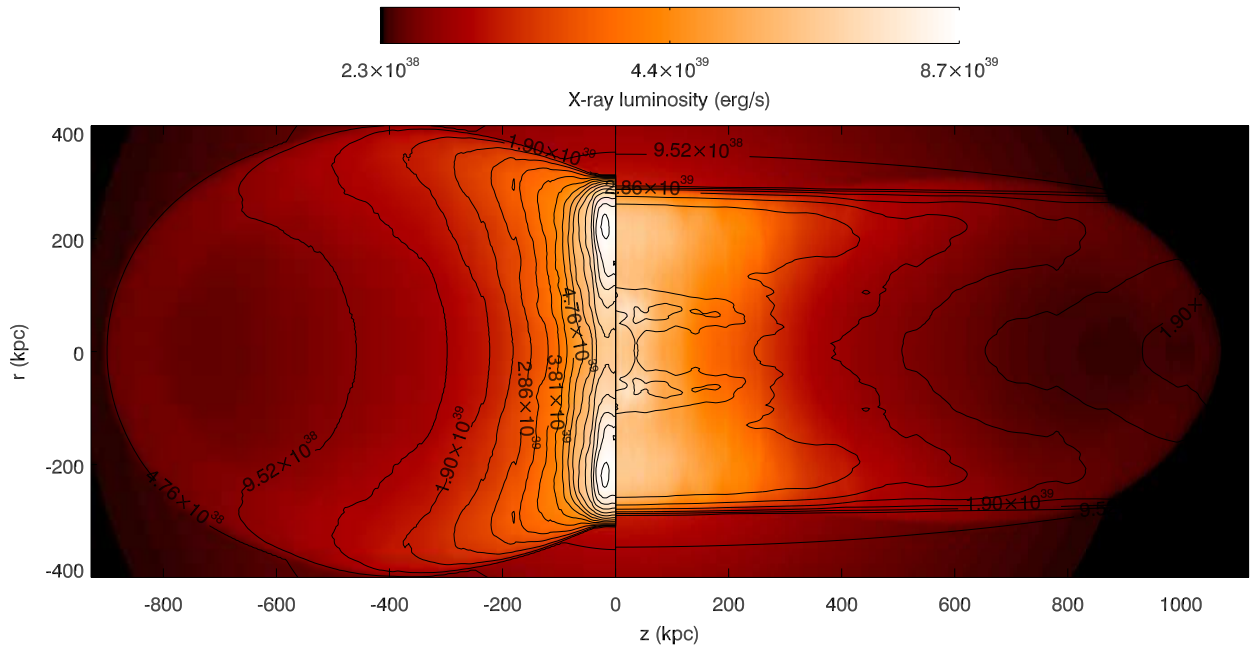}
 \caption{Synthetic X-ray luminosity map extracted from the last snapshot of simulations J46 (left) and J46n (right). The X-ray cavities produced by J46n are smaller than those produced by lighter relativistic jets. The X-ray images shown in this figure has been obtained by integrating the X-ray luminosity along a $90^\circ$ line of sight, after applying axial symmetry revolution to the two-dimensional simulation. The bolometric (integrated) luminosity is $L_{\rm x}\simeq2\times10^{45}\,{\rm erg/s}$ in both cases. The bright ridge perpendicular to the jet axis is a consequence of the lack of a detailed set-up for the host galaxy.}
 \label{fig:xrn}
 \end{figure*}

      The difference in the morphology of the cocoons of models J46n and J45b with respect to J46 and J45l, as discussed previously in this section, reflects also in their corresponding X-ray maps. Figure~\ref{fig:xrn} shows that J46n does not produce an X-ray cavity with the same morphology and size as that obtained for its relativistic counterpart. Thus, we can state that those jets cannot be baryonic dominated, and that a significant contribution of non-thermal particles or pairs to the jet flow is required, in agreement with observational analysis of multiwavelength emission in a quasar jet at parsec scales \citep{ka08}.

%
%

\subsection{On heating}
 \label{sec:agnkf}

\subsubsection{On AGN kinetic heating}

 \cite{sha11} have introduced the idea of a dual-mode AGN kinetic feedback between galactic activity and ambient media, depending on the nature of the jet. In the case of low-power jets (jet kinetic power $\leq 10^{44}\,{\rm erg/s}$), feedback can be more gentle and focused on the immediate region around the host galaxy, affecting up to $\sim$100~kpc, whereas in the case of powerful jets (jet kinetic power $> 10^{44}\,{\rm erg/s}$), the feedback can be much faster and heat, not only the surroundings of the galaxy, but also larger regions up to hundreds of kiloparsecs. In addition, the large-scale shocks detected around a number of radio-galaxies must have played a fundamental role in the heating of the gas surrounding their host galaxies \citep[e.g.,][]{sta14}. 
    
Our set of simulations, which span two orders of magnitude of jet kinetic powers, allows us to give support to this scenario.
 J44 represents a low-power jet, which is well-collimated during the active phase, due to the absence of growing helical instabilities, which are forbidden by the two-dimensional nature of the simulations, or strong recollimation shocks. The large difference in head advance velocity with the other jets limits the deposition of energy to a smaller region ($\sim 100\,{\rm kpc}$ mean radius) during the active phase. Later, the weak shock slowly expands up to $\sim 300\,{\rm kpc}$, but the hot plasma injected is only located in the central regions. These results must be scaled depending on the properties of the ambient medium \citep{ka09}: the distance travelled by jets will be smaller for denser media. Shocks have been detected even in less-powerful jets \citep{kr07,cro07}, and this is also confirmed by numerical simulations \citep{pm07,bbrp11}, so we expect that mechanical heating is active in low-power jets during the first Myrs of evolution. 
 
     In contrast, the powerful jets J45l, J45b and J46 expand rapidly and heat large regions beyond the host galaxy, up to several hundred kiloparsecs, mainly via strong shocks. During the active phase heating is focused along the privileged axis of propagation of jets, but it is rapidly isotropized when the advance speed of the head is reduced in the passive phase. 
Our results thus support  the dual-mode kinetic feedback proposed by \citet{sha11},
with powerful radio sources associated to heating by strong shocks, and low-power
radio sources associated to a more gentle heating mechanism,
involving weak shock waves and mixing.

%
\begin{figure*}
 \includegraphics[width=0.45\textwidth]{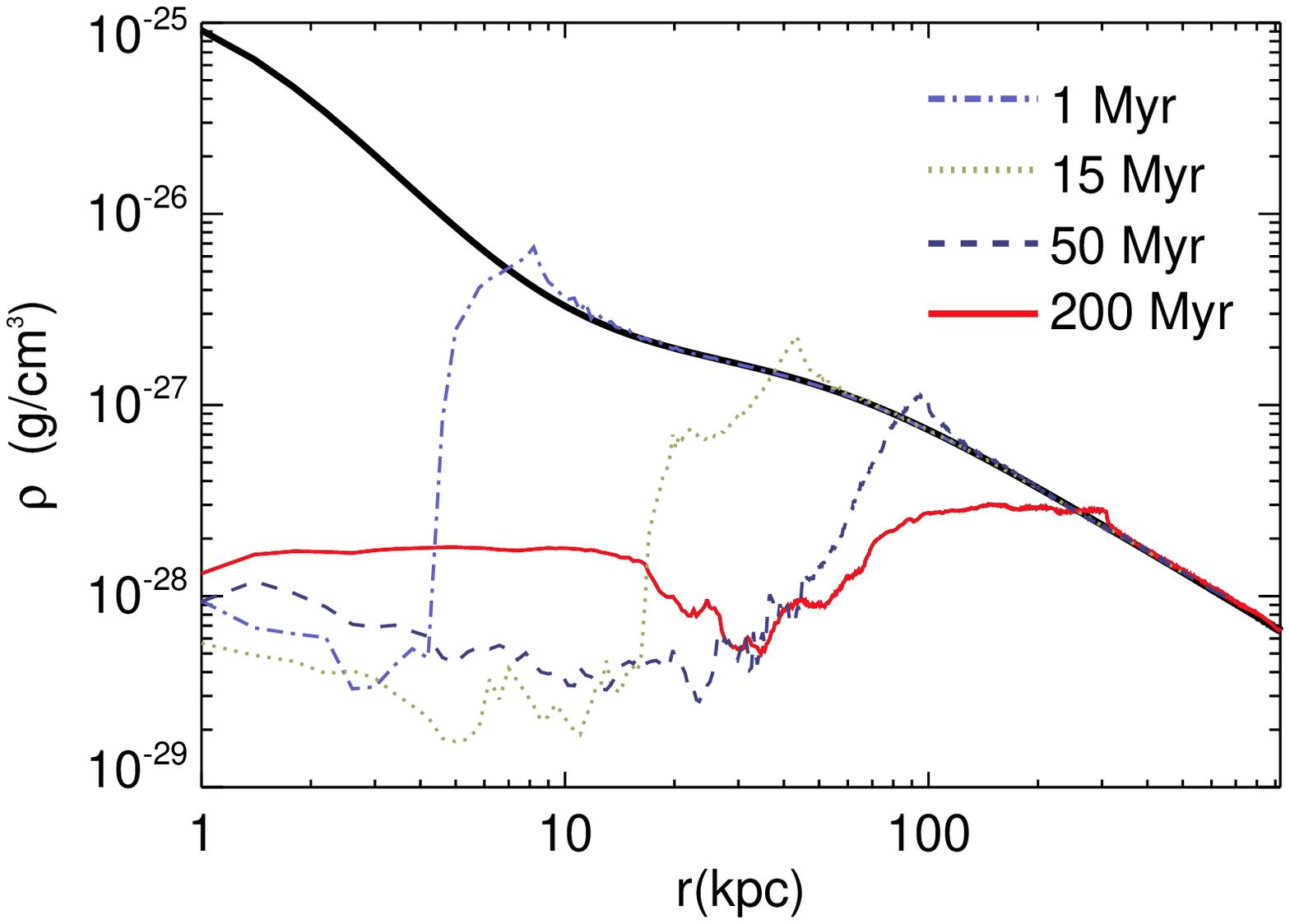}
 \includegraphics[width=0.45\textwidth]{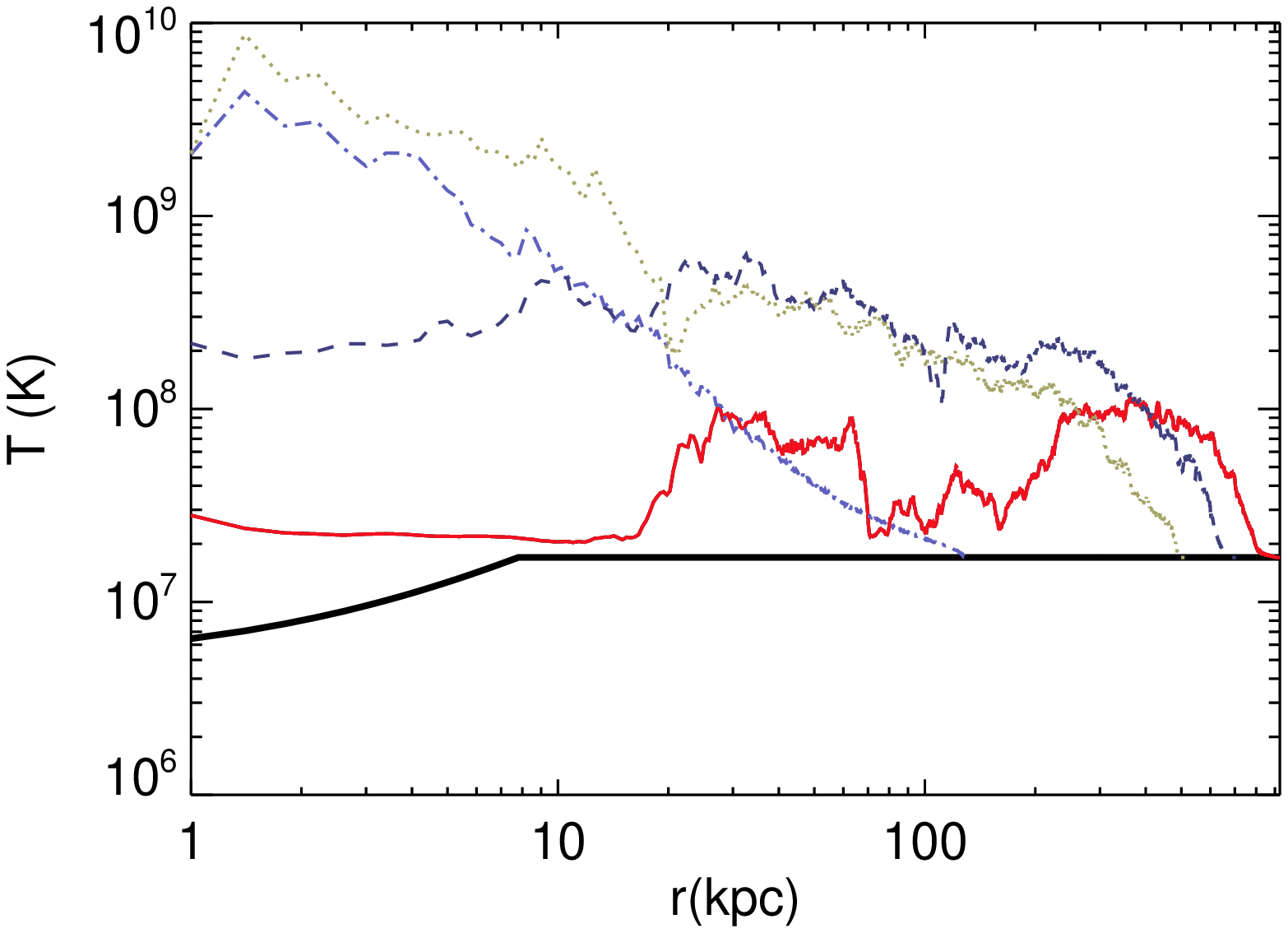}\\
 \includegraphics[width=0.45\textwidth]{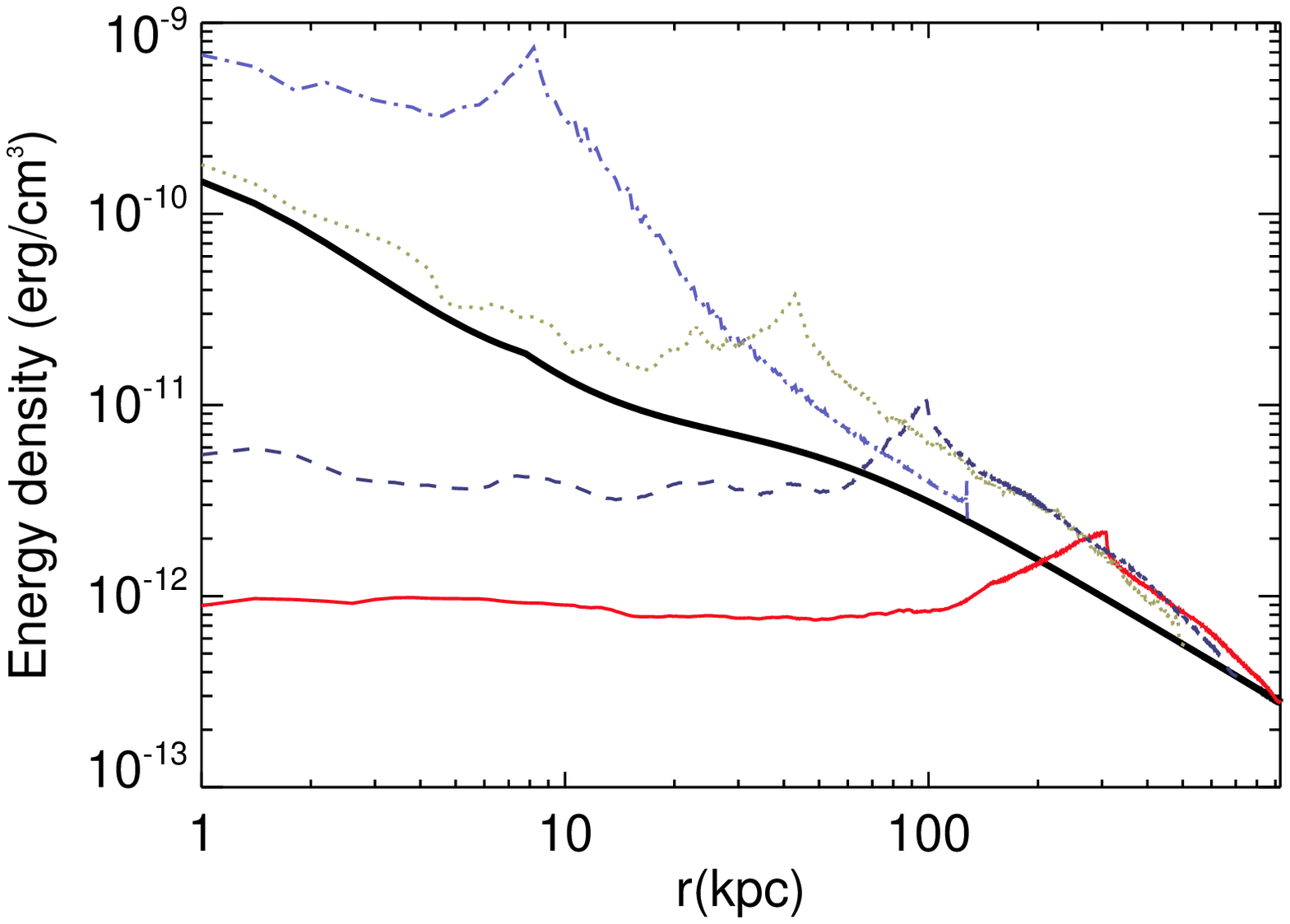}
\includegraphics[width=0.45\textwidth]{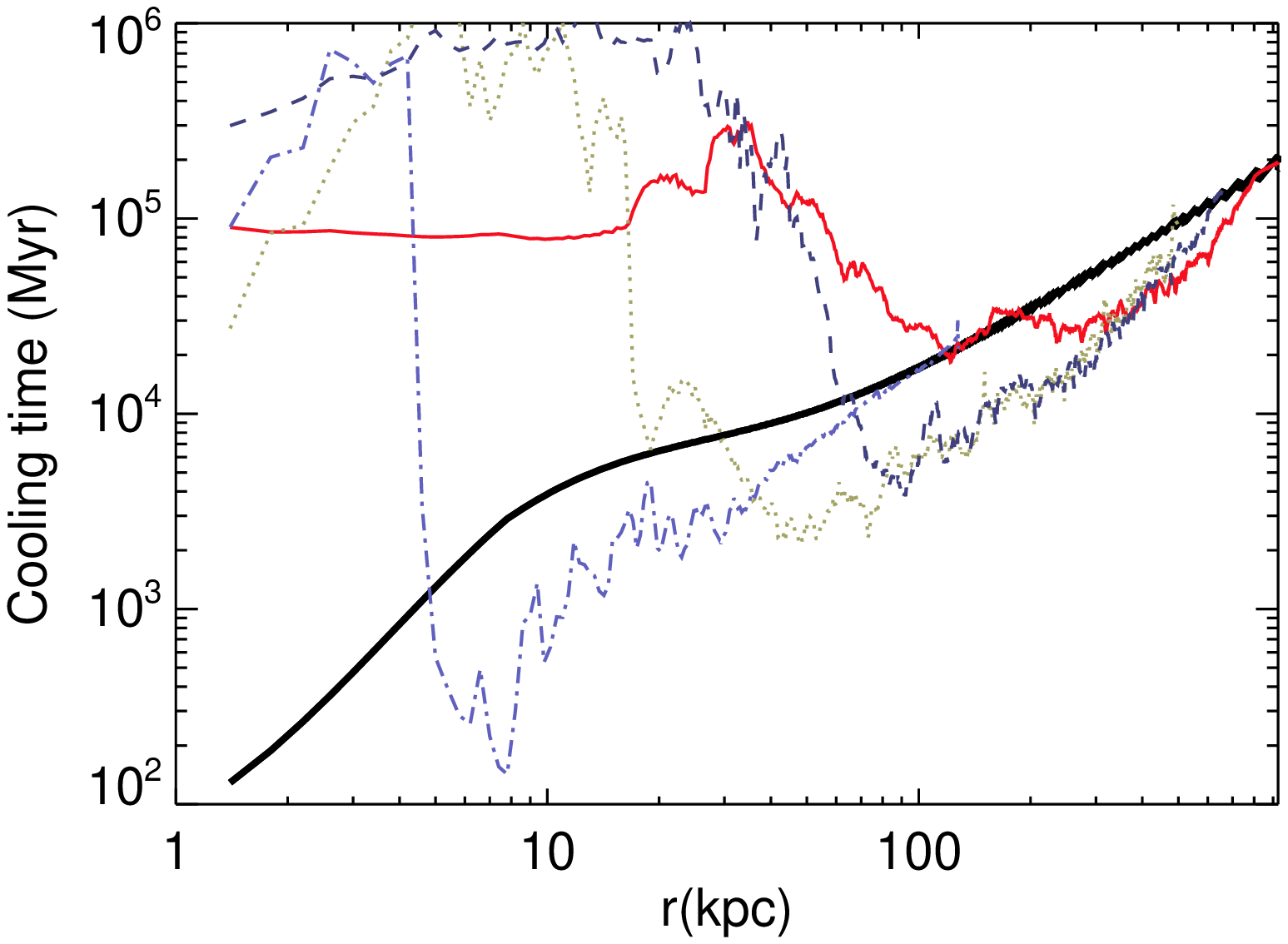}
 \caption{Mean density (top left), temperature (top-right), internal energy (bottom-left) and cooling times with distance to the galactic nucleus for J46. The blue solid line represents the mean values at $t\simeq1\,{\rm Myr}$, the green dotted line at $t\simeq15\,{\rm Myr}$, the dark-blue dashed line at $t\simeq 50\,{\rm Myr}$, and the red line at $t\simeq 200\,{\rm Myr}$. The thick, black line indicates the initial equilibrium state.}
 \label{fig:mJ46}
 \end{figure*}
%

%
\begin{figure*}
 \includegraphics[width=0.45\textwidth]{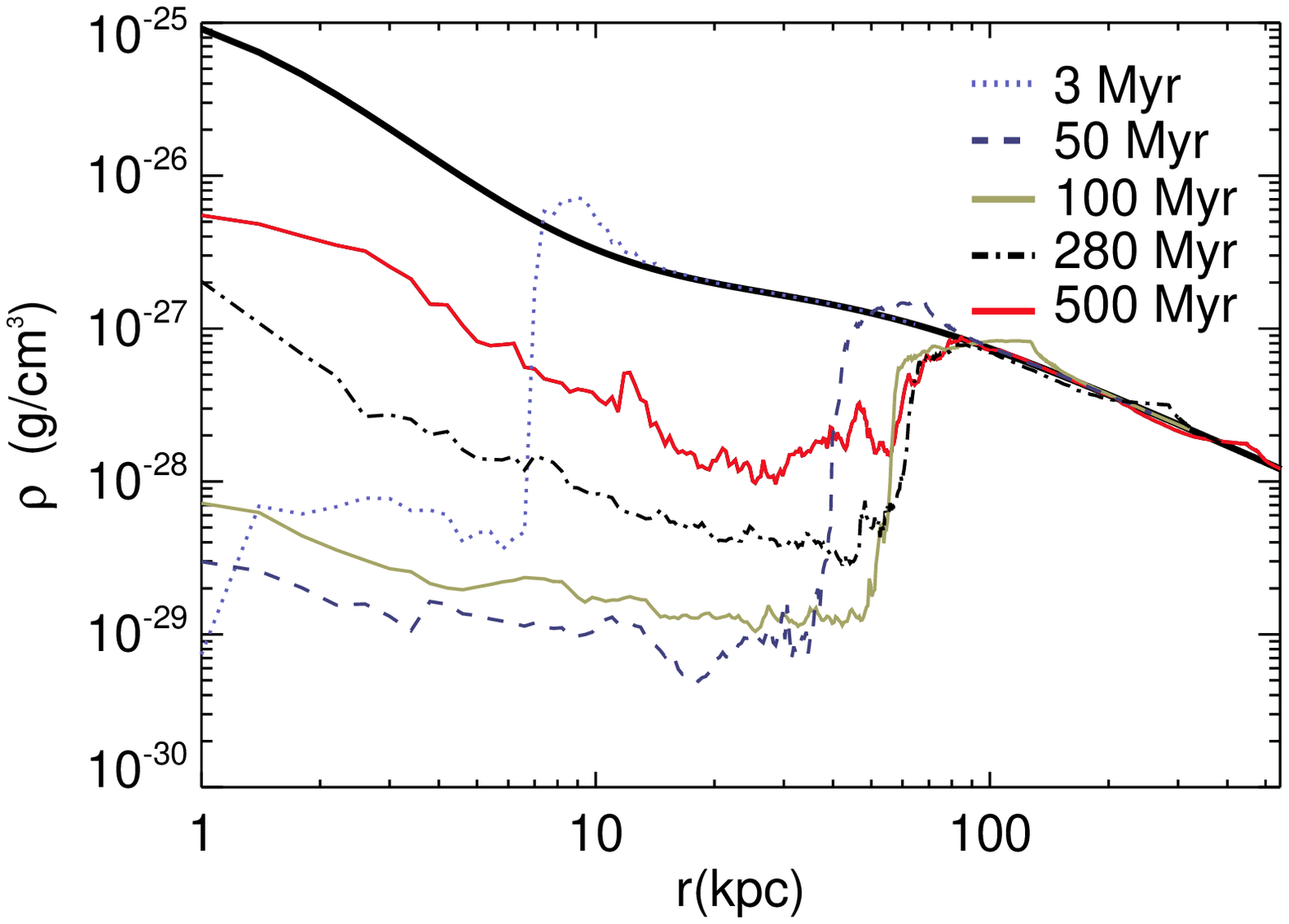}
 \includegraphics[width=0.45\textwidth]{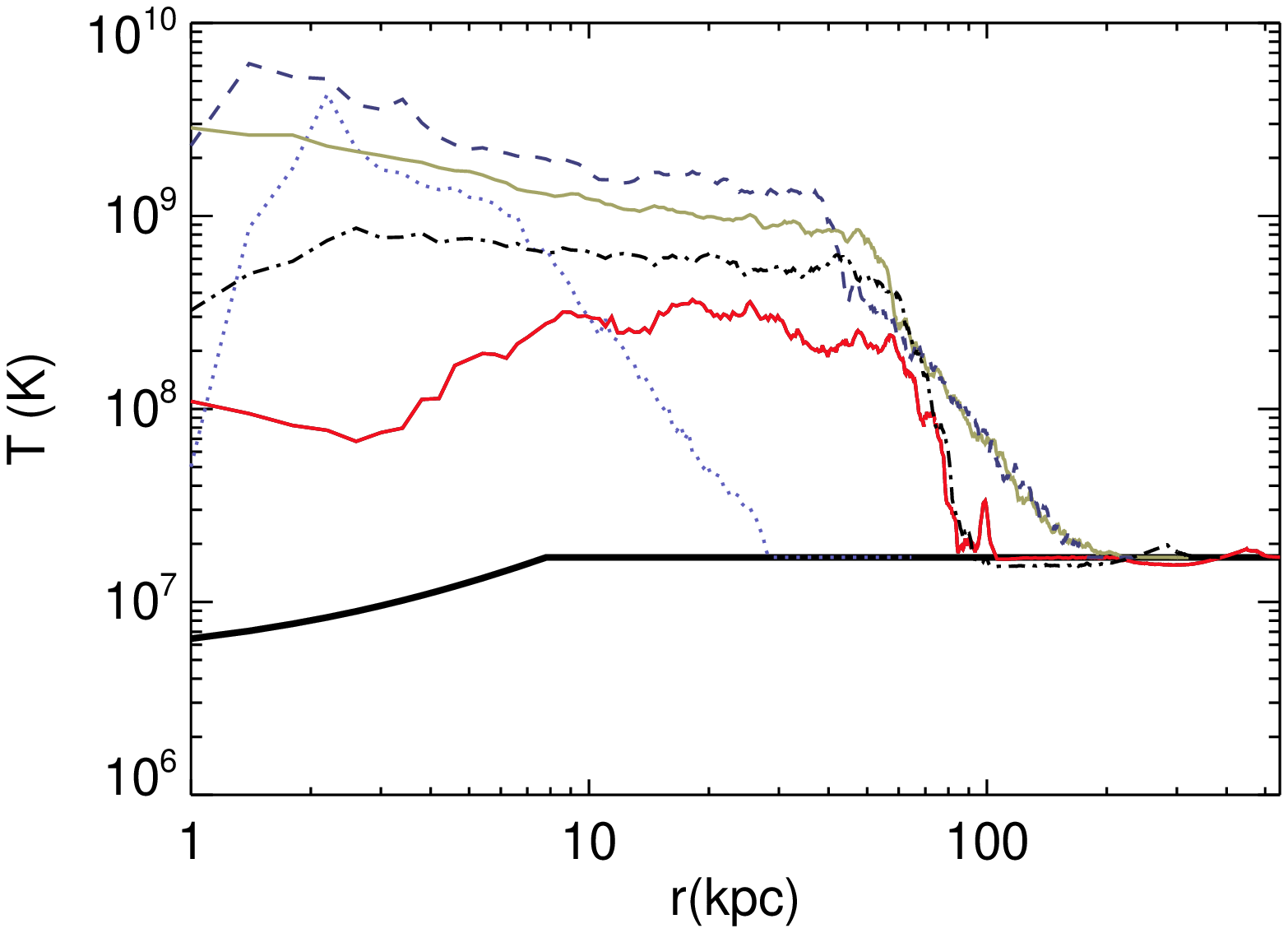}\\
 \includegraphics[width=0.45\textwidth]{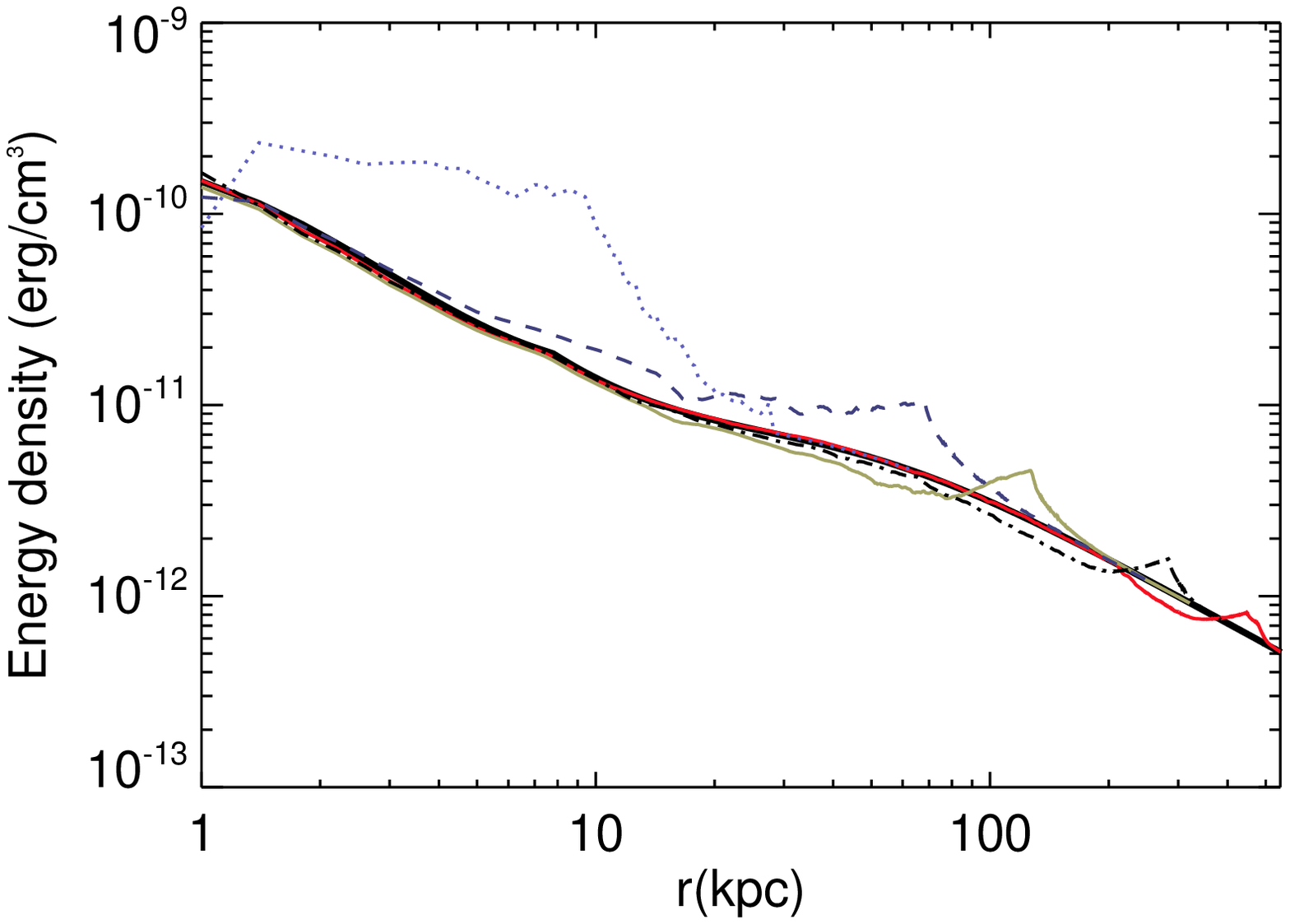}
\includegraphics[width=0.45\textwidth]{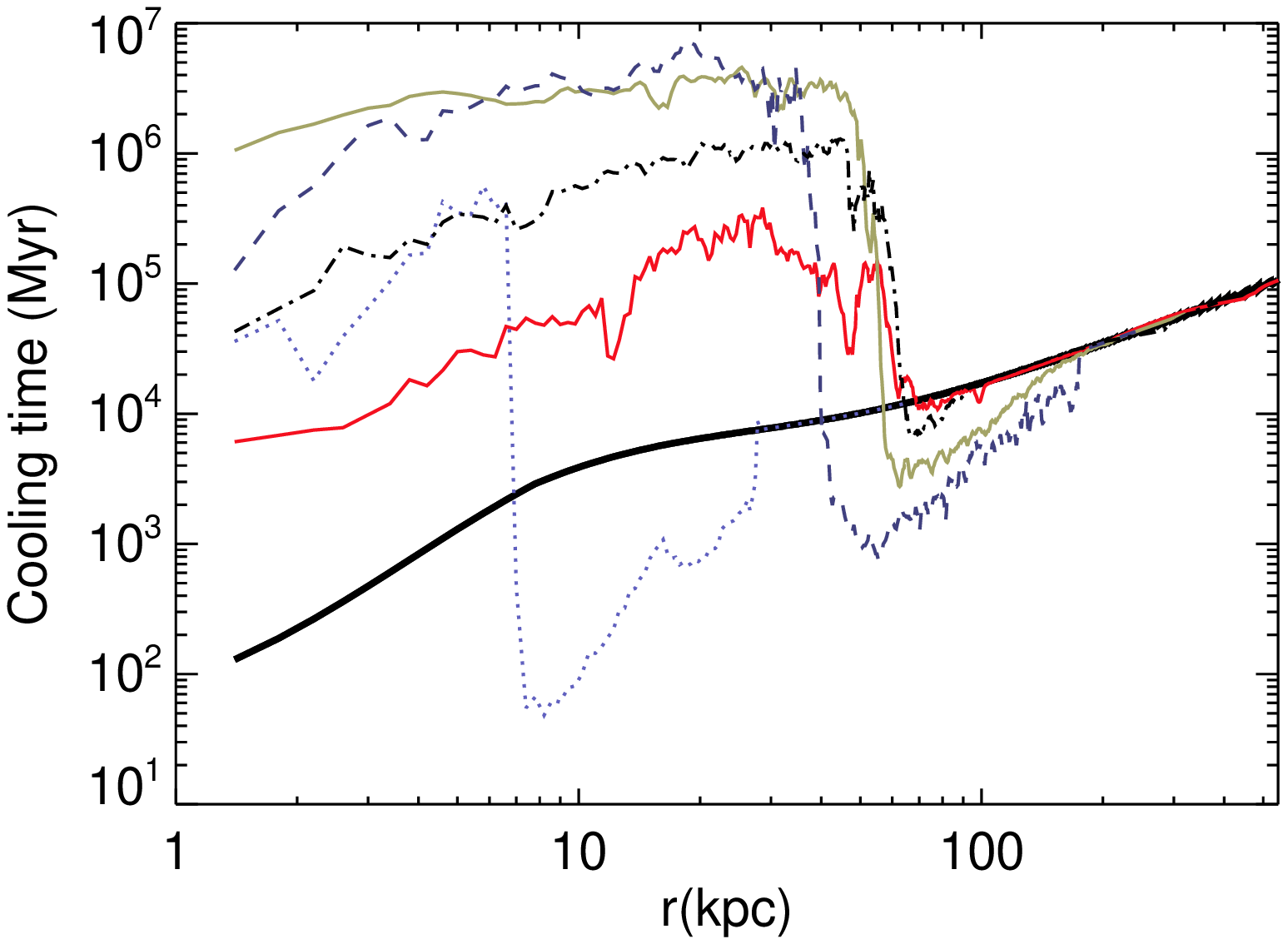}
 \caption{Mean density (top left), temperature (top-right), internal energy (bottom-left) and cooling times with distance to the galactic nucleus for J44. The blue solid line represents the mean values at $t\simeq3\,{\rm Myr}$, the dark-blue dashed line at $t\simeq 50\,{\rm Myr}$, green dotted line at $t\simeq 100\,{\rm Myr}$, the black dashed line at $t\simeq 280\,{\rm Myr}$, and the red line at $t\simeq500\,{\rm Myr}$. The thick, black line indicates the initial equilibrium state.}
 \label{fig:mJ44}
 \end{figure*}
%

%
\begin{figure*}
 \includegraphics[width=0.45\textwidth]{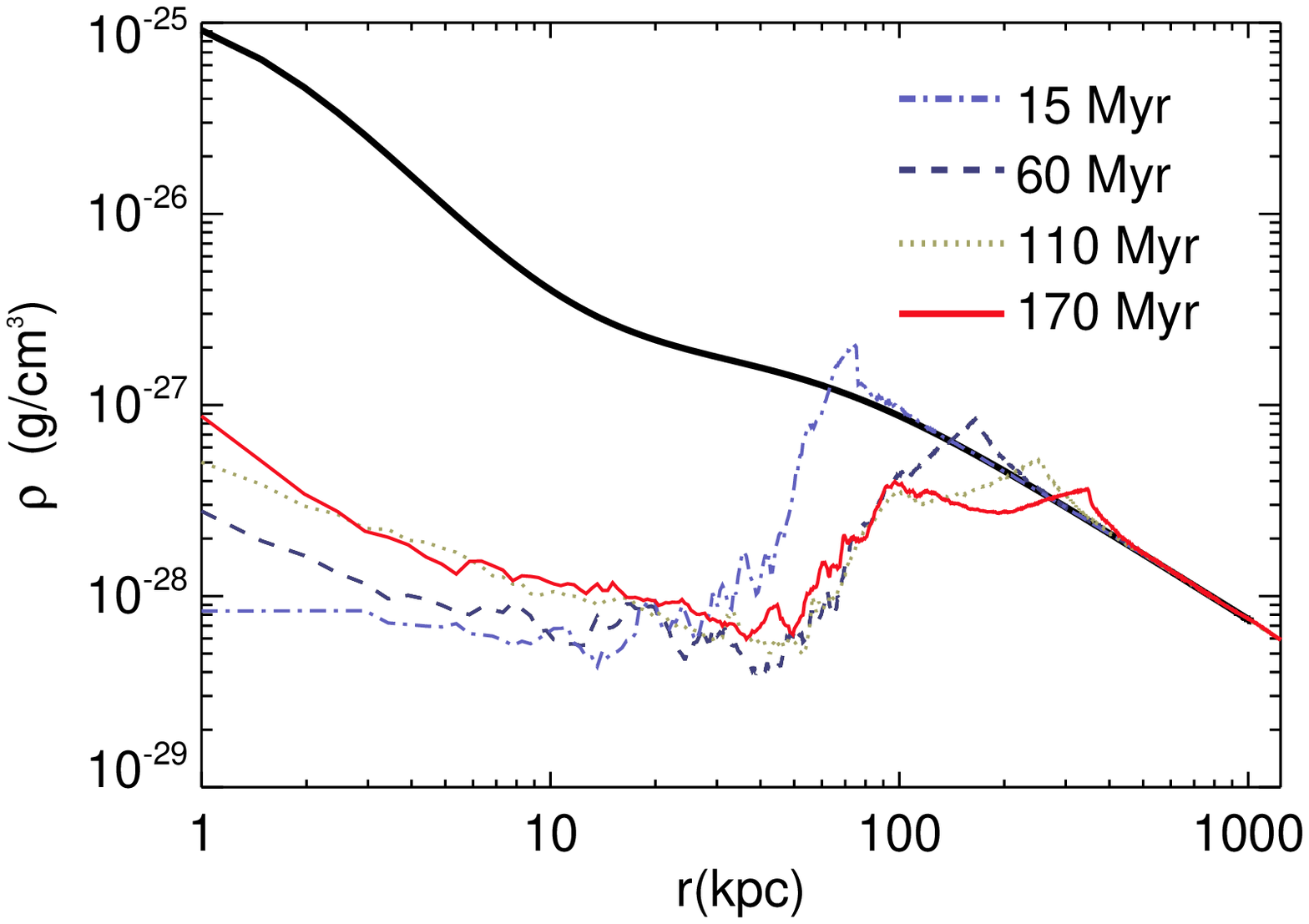}
 \includegraphics[width=0.45\textwidth]{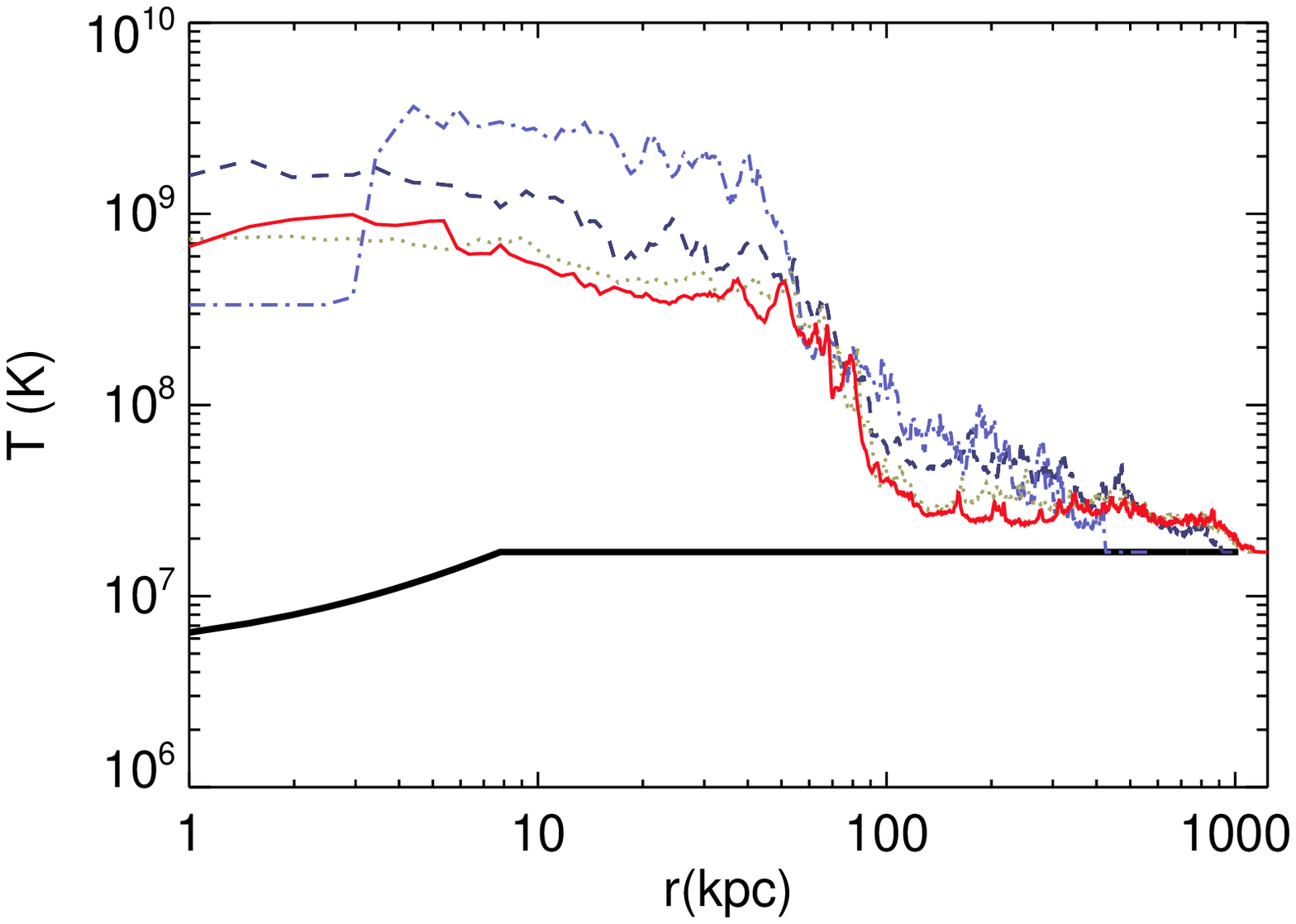}\\
 \includegraphics[width=0.45\textwidth]{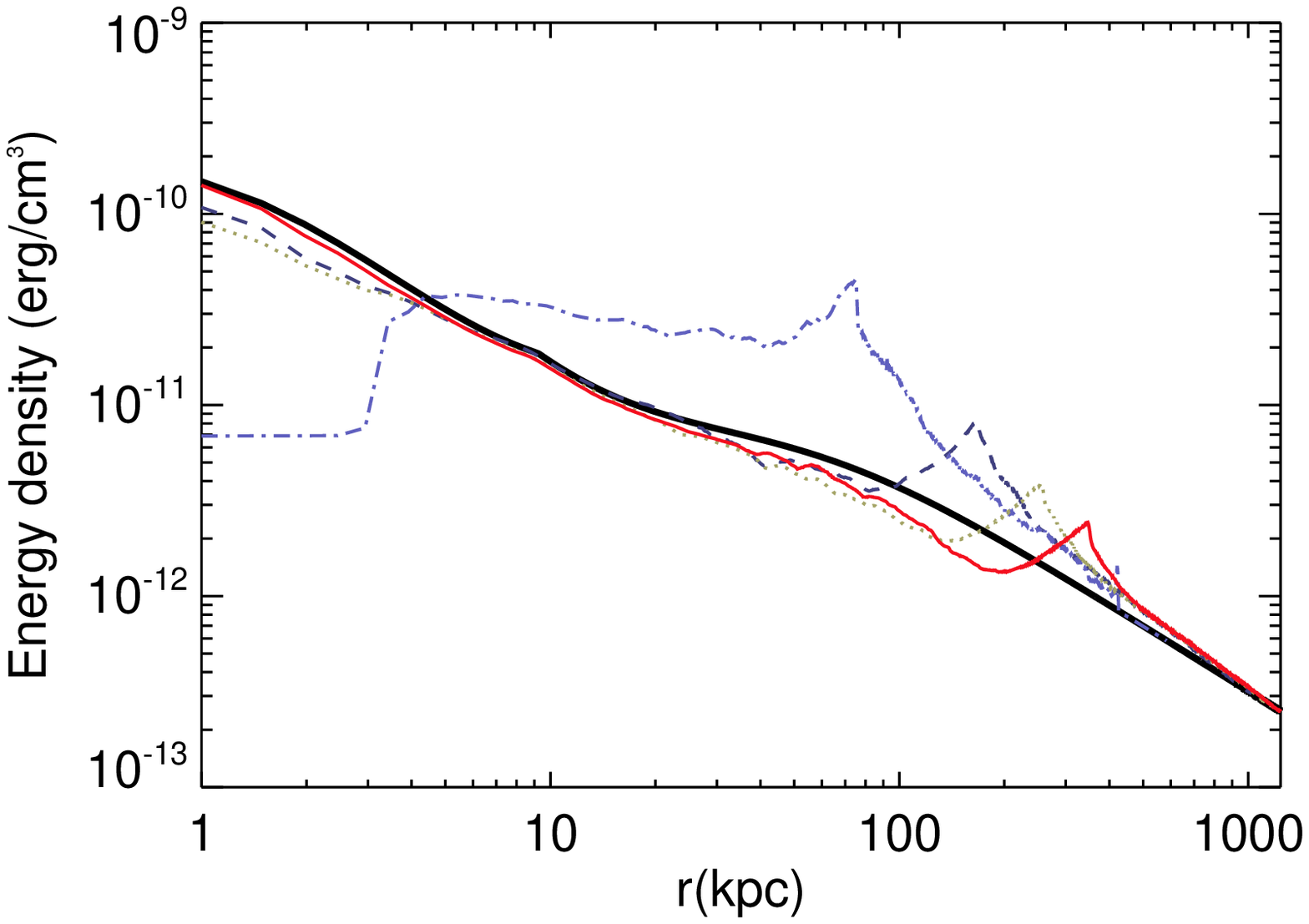}
\includegraphics[width=0.45\textwidth]{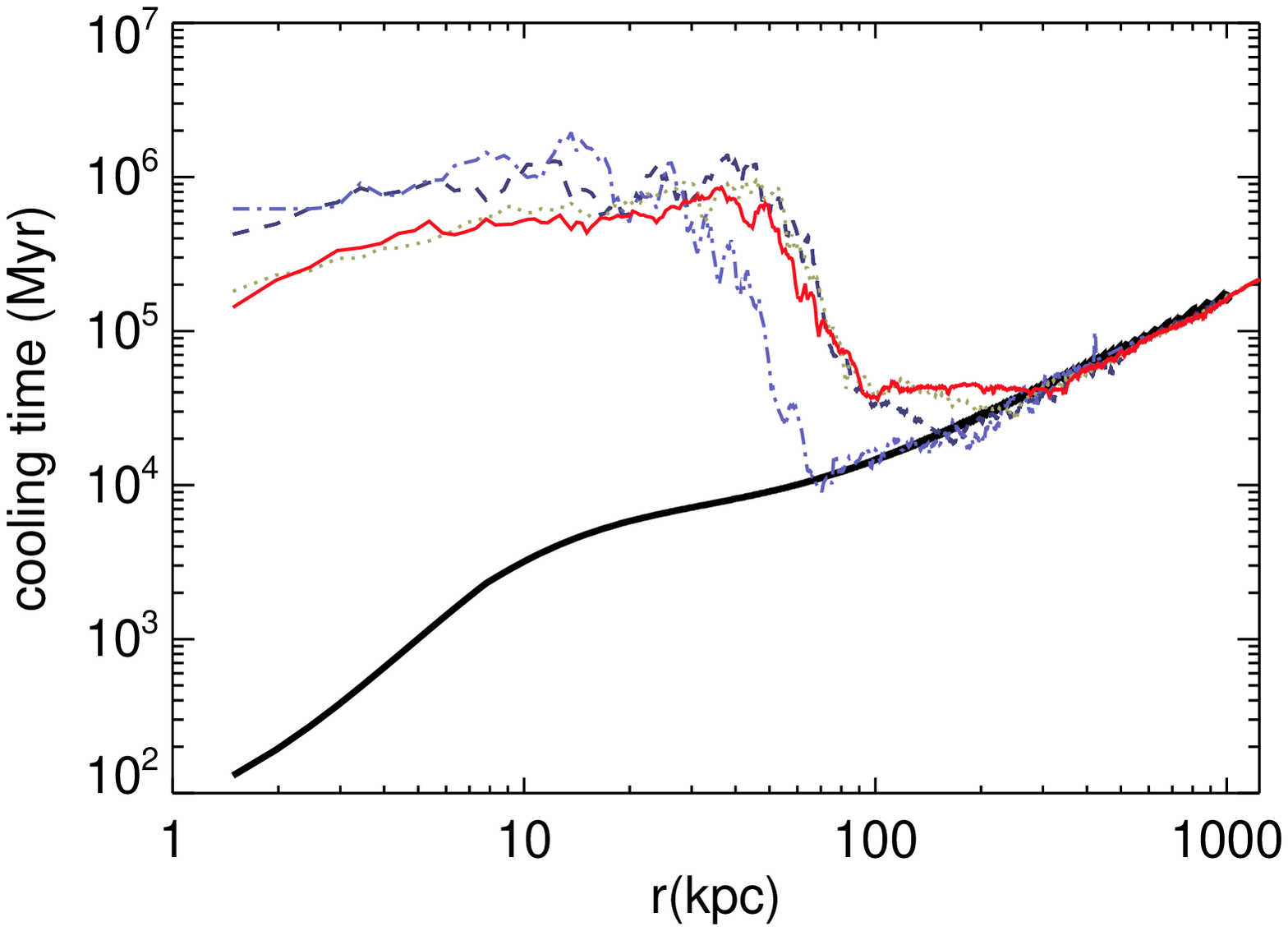}
 \caption{Mean density (top left), temperature (top-right), internal energy (bottom-left) and cooling times with distance to the galactic nucleus for J46n. The blue solid line represents the mean values at $t\simeq15\,{\rm Myr}$, the dark-blue dashed line at $t\simeq 60\,{\rm Myr}$, green dotted line at $t\simeq 110\,{\rm Myr}$, and the red solid line at $t\simeq170\,{\rm Myr}$. The thick, black line indicates the initial equilibrium state.} 
 \label{fig:mJ46n}
 \end{figure*}
%

 The different scales at which low- and high-power jets  operate are also inferred from radial profiles of the relevant quantities. Figures~\ref{fig:mJ46}, \ref{fig:mJ44} and \ref{fig:mJ46n} show the radially averaged distributions of rest-mass density, gas temperature, internal energy density and cooling times for the cases of J46, J44 and J46n, respectively, at different times during the simulated evolution. These profiles have been obtained taking into account the axisymmetric nature of the simulations, reconstructing their 3D structure, and averaging the resulting distribution in spherical shells. We would like to stress that these figures cannot be directly compared with observations as they are calculated from the simulations, taking into account the whole physical information. The thick, black lines correspond to the profiles before injection. In all cases, the region within the cocoons is emptied of gas and heated, which results in very long cooling times. In the case of J44, this region is of the order of tens of kiloparsecs, whereas in the case of J46 it extends to a hundred kiloparsecs. Cooling times are relatively short only in regions in which the gas is compressed by the shock. However, this gas is flowing out so it cannot form a cooling flow onto the galaxy. During the passive phase, entrainment of shocked gas within the formed cavity rises the density of the gas and cools it down. As a result, the cooling times decrease again, but stay still far from the original one, which is equivalent to that given in \cite{hr02} by construction. An important difference between J46, on the one hand, and J44 and J46n, on the other hand, is the change in the internal energy profile in the case of J46 with respect to the rapid recovery of the initial profile in J44 and J46n. In the absence of energetic losses, and taking into account that the energy injected by the jets in our simulations is a minor fraction of the rest-mass energy in the ambient medium, the systems tend to recover their equilibrium state, forced by the gravitational potential. Taking into account that the this potential is the same for all the simulations, the difference in the relaxation times is proportional to the ratio of injected energies. The main actor of this process is the pressure that drives the expansion of the shocks ($p=E_{\rm injected}/V$): The larger the pressure, the larger the force exerted for the expansion, and the longer the time needed for the gravity to balance this force. The difference in the total injected energy gives the difference in the relaxation times between J46 and J44. In the case of J46  the cause of the longer relaxation time is the difference in the thermal pressure as compared to J46n, where the mass of the particles represents a larger proportion of the injected energy, with respect to J46 (see top panels of Fig.~\ref{fig:PRT}).

 Our simulations thus demonstrate that powerful jets have a
strong impact on the ambient medium 
 by efficiently heating and displacing a large amount 
of the ambient gas (up to $\sim 10^{11}\,M_{\odot}$, see Paper~I).
The increase of the gas temperature and the long cooling times that we
find imply a significant reduction of the gas available for star
formation in the central galaxy, thus having a dramatic impact on galaxy
evolution \citep[see][]{sijacki07, dub13}.  Furthermore, by heating the
ICM at large distances from the central galaxies,  we expect the AGN
feedback to have important implications also on the evolution of
satellite galaxies, in agreement with the observational finding of
\cite{sha11}.

\subsubsection{Cool-core structure} 

\begin{figure}
 \includegraphics[width=0.45\textwidth]{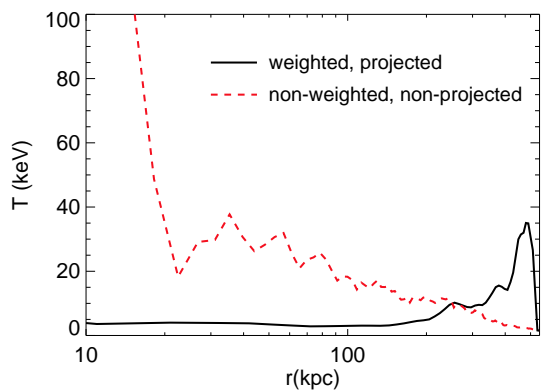}
 \caption{Radial profile of the projected, luminosity-weighted temperature for J46 at the end of the active phase ($t\simeq16\,{\rm Myr}$, black-solid line), and radial profile obtained directly from the gas temperature as given by the simulations, i.e., non-projected and non-weighted (red-dashed line).}
 \label{fig:lwt}
 \end{figure}

 The profiles in Figs.~\ref{fig:mJ46}, \ref{fig:mJ44} and \ref{fig:mJ46n} (in which the gas temperature does not show any cool central region) cannot be directly compared to observations because they are obtained from the $r-z$ simulation plane in spherical shells from the origin. In order to compare our results with observational estimates, the projected values should be calculated. We have calculated the projected, luminosity-weighted temperature by 1) computing a three-dimensional distribution of thermal X-ray luminosity-weighted temperature following the axisymmetric nature of our simulations, and 2) projecting along the line of sight for each pixel (at a viewing angle of 90$^\circ$). Figure~\ref{fig:lwt} shows the radial profile of this projected, luminosity-weighted temperature for J46 at the end of the active phase ($t\simeq16\,{\rm Myr}$), and the profile obtained directly from the gas temperature as given by the simulations, i.e., non-projected, and also shown in Fig.~\ref{fig:mJ46}. When deriving the projected, weighted temperature profile, the contribution of external cells with lower-temperatures, but much brighter in thermal radiation, decreases the mean value. In addition, it makes the front shock more visible in the plot for the same reason. On the contrary, in the case of the non-projected profile, the inner region is dominated by the hot cells within the cocoon (Figs.~\ref{fig:mJ46}, \ref{fig:mJ44} and \ref{fig:mJ46n}). Although we only show the temperature profiles for J46, the change in the temperature gradient is observed for all the simulations. 
    
    As a result the profile of projected, luminosity-weighted temperature (black-solid line in the figure) shows a flat distribution in the inner region and an increase towards the location of the bow-shock. This distribution resembles observed cool-core profiles of luminosity-weighted temperatures in different clusters that host powerful radio sources \citep[see, e.g.,][]{mc05,si09a,wo08,git10,sta14}. 
             
   Therefore, taking into account that 1) the luminosity-weighted temperature is necessarily biased towards the values in the most luminous regions, and that 2) this reduces the weighting of the volume filled by the hot, dilute plasma injected by relativistic jets in AGN, it is not possible to neglect this kind of mechanical feedback even in clusters that show cool-cores. Cooling effects, which were not included in our simulations, are not expected to play any role in this result, because the longest cooling times show up in the central regions. Moreover those cooling times are larger than the total simulated time (Figs.~\ref{fig:mJ46} and \ref{fig:mJ44}).

%
%

\subsection{Comparison with other works, caveats and future work}
\label{comp}
 We present here the first relativistic simulations aimed to study the long-term evolution of powerful AGN jets in the context of feedback. Recent work by \citet{wb11,wbu12} shows a study of the influence of relativistic outflows in the host galaxy for the first $10^5\,{\rm yr}$ of evolution and its importance in quenching star formation within the galaxy. Although our simulations are focused on largest scales and include a less detailed description of the ambient medium, our results point in the same direction, as shown by the large amount of gas that is displaced by shocks. We also performed a non-relativistic numerical simulation with the aim to compare the results obtained, not only with the relativistic simulation J46, but also with previous works that use this approach. In this section we point out several differences with other works that can explain the different results obtained. 
 
    Among previous works, we can distinguish between simulations in which the injection of the jet is imposed as a poorly-collimated or collimated flow. In the former, the conservation of momentum flux forces the outflow propagation velocity to decrease with distance, so that any shock wave that can be initially generated tends to weaken and disappear, becoming a sound wave \citep[e.g.,][]{vr06,vr07,ss09}. On the contrary, our results are comparable to those derived in the case of collimated (and continuous) injection: these works stress the importance of shocks in the interaction between jets and the heating of the ambient medium \citep[e.g.,][]{om04,za05,br07,oj10,ads10,ga12,cie13}. In other works in which the injection of the jet is collimated \citep[e.g.,][]{ct07,gas11a,gas11b} the flow has very low velocities ($10^3\,-\,10^4\,\,{\rm km\,s^{-1}}$) and in some cases, it is not continuous \citep[e.g.,][]{gas11a}, triggering very different evolution and results to those presented in this paper. Another difference is the presence of a galactic plus a cluster component in the ambient medium in our simulations as compared to a single cluster component in previous works. In addition, the asymptotic $\beta$-parameter of ambient medium density and pressure profiles, $\beta_{\rm atm,g}$ in Eq.~(\ref{next}) is typically $0.5$ in the aforementioned works, whereas it is $0.38$ in our case, with the steeper gradient favoring buoyant motion.
  
   \cite{mit09} pointed out that most clusters preserve the cool-core structure and do not invert their temperature gradients. However, the radio properties of most of the studied clusters show that they mainly correspond to FRI jets, in terms of the luminosity division given by \cite{gc01b}. This makes their results difficult to compare with our simulations, which are biased towards high-power sources (J44 lies at the high-power end of typical FRI jet power). Based on this observational result,\cite{gas11a,gas11b} showed that a slow, massive outflow with irregular injection power could fulfill the requirement of preserving a cool-core, even with the injection of large power outflows, and suggested that this could be a plausible mechanical feedback mode. However, our simulations show that relativistic jets in powerful radio-galaxies can also represent a fast and efficient heating mechanism reaching hundreds of kiloparsecs. We have also shown that this result is still compatible with the presence of observed cool-cores.
   
    The AGN jets are relativistic or mildly relativistic up to hundreds of kiloparsecs \citep[see, e.g., jet/counter-jet brightness asymmetries,][]{bri94}. In the case of FRII jets, they keep a high degree of collimation and do not show any hints of a wider, slower and massive surrounding flow, even in the case of FRI's at those scales. Following the comparison that we have performed here between jets J46 and J46n, we can state that although those previous works can be applicable to the observed galactic winds in some active galaxies, they cannot be taken as simulations of the interaction of relativistic flows with the ambient medium nor derive any implications regarding the heating of the ICM by the impact of relativistic flows or their influence on the presence or absence of cooling flows in such scenarios. It is possible that owing to the lower efficiency of slow and massive galactic outflows, the temperature gradient would remain basically unchanged in that case, but these outflows must have lower powers or shorter active periods than relativistic jets. Otherwise they also generate strong shocks and change the gas temperature profile, as shown by our simulation J46n (see Fig.~ \ref{fig:mJ46n}). In addition, massive, slow flows are only observed at distances that range from several kiloparsecs to tens of kiloparsecs at most \citep[e.g.,][]{mo05,mo07,ho08,nes08,gui12,mo13} and appear typically associated with a faster radio-jet with much larger sizes, so they could well be a consequence of the action of the relativistic jet on the ambient medium, more than a main actor of the whole process \citep[e.g.,][]{nes08,gui12,mo13}.

     Our two-dimensional simulations prohibit the development of antisymmetric unstable waves such as helical Kelvin-Helmholtz modes and favor jet collimation and fast propagation. \cite{bmk08} and \cite{abm12} computed the maximum lifetime of FRII radio sources by comparing the results from the model of evolution by \cite{ka97} and \cite{kda97} and mock catalogs of known jet, lobe and ambient medium properties. They obtained a typical lifetime of $1.5\times10^7\,{\rm yr}$ in the case of galaxy groups\footnote{This number underestimates the age in a factor five, as discussed in \cite{abm12}.} and of $1.9\times10^8\,{\rm yr}$ in the case of galaxy clusters. Our simulated jets propagate through hundreds of kiloparsecs within a few tens of Myr, which represents a faster propagation, owing to lower ambient medium density and axisymmetry. It is important to remark that the ambient medium used in our work corresponds to that of a radiogalaxy with an old jet \citep[3C31, with an estimated age larger than $10^8\,{\rm yr}$,][]{pm07} that could have changed the original properties. Future work will include fully three-dimensional simulations and careful modeling of the ambient medium to study, on the one hand, the influence of the ambient medium properties in the process of jet, shock and cavity evolution, and, on the other hand, the effect of the interaction on the evolution of clusters hosting active galaxies and relativistic outflows.

\section{Summary and conclusions}
\label{sum}

 We have performed the longest simulations of axisymmetric two-dimensional jets evolving in a realistic pressure and density profiles. Our results show the large energetic efficiency of ambient medium heating by AGN jets, and the long-lasting existence of weak shocks that can dominate the cavity expansion until hundreds of millions of years in the case of large-power, collimated outflows. We have shown that this evolution follows the eBC model \citep[][and Paper I]{pm07} with high accuracy, and that buoyant motion is still not the main driver of the evolution of cavities by the end of our simulations. The deviation from the Sedov phase by the end of the simulations is interpreted as a phase in which the evolution is dominated by weak-shocks and possible transition to transonic speeds. This behaviour is observed mainly in the case of the low-power jets, and further evolution could lead to buoyant motion. A simulation of slow, massive, non-relativistic jet with usual properties used in the literature, has also been performed. The increase in the ambient medium entropy caused by the non-relativistic jet involves a smaller region and the work done in generating the cavity is an order of magnitude smaller than in the case of its relativistic counterpart. In addition, we have shown that leptonic dominated jets end up with more similar aspect ratios to observed large-scale lobes than baryonic jets like the non-relativistic one. The simulation of the most powerful (leptonic) jet recovers the observed gross morphology of clusters embedding jets that are clearly associated with detected shocks: Hercules~A \citep{nu05}, Hydra~A \citep{si09b}, MS0735.6+7421 \citep{mc05}, HCG~62 \citep{git10}, 3C~444 \citep{cro11} or PKS B1358-113 \citep{sta14}. Therefore, we conclude that an important fraction of the jet's energetic budget has to be in the form of internal energy to explain the morphology of the X-ray cavities. In general, it is assumed that the cavities and the surrounding ambient medium are in pressure equilibrium, on the basis of pressure estimates that are obtained from the non-thermal population in the radio-lobes. However, our simulations show that ongoing mixing at the contact discontinuity between the shocked jet and ambient gas introduces an important amount of cold gas in the cavity, resulting in cavity overpressure and the presence of weak shocks well after the jet active phase. 
 
   Following our results, we claim that there is not a single kinetic feedback mode from AGN, but that the main way in which powerful radio galaxies heat their environment strongly depends on the properties of the outflow that is generated. However, given the ample spectrum of outflow energies and variety of manifestations of galactic activity, the mechanism may vary from pure heating by slow mixing, heat transfer, local heating by slow and massive outflows or radiative heating \citep[see, e.g.,][]{mn07,fb12} to heating by strong shocks. Our results show that in the case of radio galaxies with powers $\gtrsim 10^{44}\,{\rm erg/s}$, the main heating mechanism is mechanical heating by shocks, which also seems to be enough to stop cooling flows. This is consistent with the dual-mode AGN kinetic feedback suggested by \cite{sha11}. We have shown that luminosity-weighted temperature profiles could give cool-core structures even in the sources in which shocks have been detected. Our results show that a hot inner region with long cooling times could be hidden even in the case of mechanical heating by powerful AGN jets, as the cool-core structure is recovered by the luminosity-weighted temperature also in that case. This hot, thermal component of dilute gas with temperatures around $10^9~{\rm K}$ could be detected with ALMA via the Sunyaev-Zeldovich effect \citep{co13}. Finally, we conclude that the amount of gas that is expelled from the galaxy by strong shocks should result in a drastic quenching of star-formation in those galaxies and thus, in their reddening.  
 
  Future work should exploit three-dimensional simulations, including cooling terms in the equations and an improved initial set up, to account for a more detailed description of the ambient medium.

\section*{Acknowledgments}

The simulations presented in this paper were carried out at the Barcelona Supercomputing Centre (\emph{Mare Nostrum}) using resources from the Spanish Supercomputing Network (RES). Analysis of the simulations and support calculations were performed at the computing facilities of the \emph{Centre de C\`alcul} of the University of Val\`encia, including \emph{Tirant}, \emph{Llu\'{\i}s Vives}, and \emph{CERCA2}. The authors acknowledge financial support by the Spanish ``Ministerio de Ciencia e Innovaci\'on'' (MICINN) grants AYA2010-21322-C03-01, AYA2010-21097-C03-01 and 
CONSOLIDER2007-00050, and by the ``Generalitat Valenciana'' grant ``PROMETEO-2009-103''. The authors thank the anonymous referee for the useful and interesting comments.

\end{document}